\documentclass[11pt,a4paper]{article}
\usepackage{graphicx} % Required for inserting images
\usepackage{amsfonts,amsmath,amssymb} % Mathematical tools
\usepackage{float} %Figure placement
\usepackage[margin=2cm]{geometry} %Margin dimensions
\usepackage{longtable} %Multi-page tables
\usepackage{color,url}
\usepackage{colortbl}
\usepackage{multirow}
\usepackage{verbatim}
\usepackage[utf8]{inputenc}
\usepackage{booktabs}
\usepackage{upgreek}
\usepackage[colorlinks=true
,urlcolor=blue
,anchorcolor=blue
,citecolor=blue
,filecolor=blue
,linkcolor=blue
,menucolor=blue
,linktocpage=true
,pdfproducer=medialab
,pdfa=true,
pdfauthor={Jorge Alda, Jacobo Asorey, Alejandro Mir, Siannah Peñaranda},
pdftitle={Physically Consistent Parameter Inference: Transparent Machine Learning Emulation in High Energy Physics and Cosmology}
]{hyperref}
\usepackage{cite}

\usepackage{epsfig,psfrag,rotating,soul}
\usepackage{rotfloat}
\usepackage{epsf}
\usepackage{graphics}
\usepackage{xcolor}
\usepackage{graphicx}
\usepackage{bm}

\begin{document}
\thispagestyle{empty}

\vspace*{2.5cm}

\vspace{0.5cm}

\begin{center}

\begin{Large}
\textbf{\textsc{Physically Consistent Parameter Inference: Transparent Machine Learning Emulation in High Energy Physics and Cosmology}}
\end{Large}

\vspace{1cm}

{\sc
Jorge Alda$^{a,b}$%
\footnote{\tt \href{mailto:jorge.aldagallo@unipd.it}{jorge.aldagallo@unipd.it}}%
,  Jacobo Asorey$^{b,c}$%
\footnote{\tt \href{mailto:jasorey@unizar.es}{jasorey@unizar.es}}%
, Alejandro Mir$^{b,c}$%
\footnote{\tt \href{mailto:amir@unizar.es}{amir@unizar.es}}%
, Siannah Pe{\~n}aranda$^{b,c}$%
\footnote{\tt \href{mailto:siannah@unizar.es}{siannah@unizar.es}}%
}

\vspace*{.7cm}

{\sl
$^a$Dipartimento di Fisica e Astronomia ``Galileo Galilei'', Universit\`a degli Studi di Padova \& INFN Sezione di Padova, via Marzolo 8, 35131 Padova, Italy

\vspace*{0.1cm}

$^b$Centro de Astropart\'iculas y F\'isica de Altas Energ\'ias (CAPA), Pedro Cerbuna 12 50009 Zaragoza, Spain
\vspace*{0.1cm}

$^c$Departamento de F{\'\i}sica Te{\'o}rica, Facultad de Ciencias,\\
Universidad de Zaragoza, Pedro Cerbuna 12,  E-50009 Zaragoza, Spain

}

\end{center}

\vspace*{0.1cm}

\begin{abstract}
\noindent

Global fits in high energy physics and cosmology often face the challenge of exploring high-dimensional parameter spaces with computationally expensive or topologically complex likelihood functions. In this work, we present a Machine Learning framework designed to emulate complex, often non-Gaussian likelihood landscapes using gradient-boosted regression trees (XGBoost). We discuss the advantages of the Machine Learning approach in terms of computational efficiency and the resolution of confidence regions, particularly in scenarios with complex correlations or ``curved'' degeneracies. We validate this methodology by applying it to a recent analysis on flavour anomalies in semileptonic $B$ meson decays and discussing the adaptability of this framework to other phenomenological systems, such as axion-like particles or cosmology global fits. Finally, we utilise SHAP (Shapley Additive exPlanations) values to provide a transparent analysis of feature importance, ensuring that the Machine Learning predictions remain physically interpretable and consistent with the underlying physics.

\end{abstract}

\def\thefootnote{\arabic{footnote}}
\setcounter{page}{0}
\setcounter{footnote}{0}

\newpage

\section{Introduction}

Despite recent progress, the exploration of highly non-Gaussian likelihoods remains computationally challenging, motivating the development of fast and interpretable surrogate models. Contemporary analyses in particle physics and cosmology often involve exploring high-dimensional parameter spaces that are constrained by hundreds of correlated measurements. Examples include flavour anomalies in semileptonic $B$ meson decays and constraints on light new particles. The resulting likelihood functions frequently exhibit intricate structures, such as strong parameter correlations, non-Gaussian behavior, multiple local minima, and elongated degeneracy directions. At the same time, evaluating the theoretical predictions required for these likelihoods often requires computationally intensive calculations, rendering exhaustive parameter scans impractical.

Traditional statistical approaches rely on repeated evaluations of the likelihood function using Monte Carlo sampling techniques, such as Markov chain Monte Carlo (MCMC) or nested sampling. However, the computational cost of these approaches increases rapidly with the dimensionality and complexity of the problem. This has motivated increasing interest in Machine Learning (ML) techniques that can emulate expensive likelihood functions while preserving the statistical information necessary for robust inference. 

Combining global fits with interpretable ML techniques provides a powerful framework for disentangling complex correlations among observables, identifying the most informative measurements, and improving the physical interpretability of New Physics (NP) constraints in the phenomenology of particle physics and cosmology. Gradient-boosted decision trees have emerged as a particularly attractive ML algorithm due to their flexibility, robustness, and ability to accurately capture nonlinear structures.

An important challenge in this type of analysis is understanding the relationship between observables and model parameters. SHAP values provide a systematic approach to quantifying the contribution of individual features to model predictions. This enables transparent and physically meaningful interpretations of ML models.

In this work, we present a unified and interpretable ML framework for global fits in high energy physics and cosmology, in which an XGBoost gradient-boosted regression ensemble is trained to emulate the (generally non-Gaussian) log-likelihood, while SHAP values keep the resulting inference transparent and physically meaningful. Rather than tailoring the method to a single problem, we demonstrate its generality across three representative global-fit applications of increasing dimensionality and complexity, spanning both particle physics and cosmology. These include the flavour anomalies in semileptonic $B$ meson decays, light and feebly-interacting new states (axion-like particles addressing the Belle~II $B^+\to K^+\nu\bar\nu$ excess), and dynamical dark-energy fits that drive contemporary precision cosmology. In all three cases, the surrogate reproduces the exact likelihood to high accuracy and accelerates parameter-space exploration by several orders of magnitude. Furthermore, SHAP analysis recovers the physically expected hierarchy of observables and parameters, ensuring that the gain in computational efficiency does not come at the expense of interpretability.

The paper is organised as follows. Section~\ref{sec:ml} introduces the ML algorithm at the heart of the framework: the construction of the XGBoost surrogate of the $\chi^2$ function, the underlying statistical formalism (likelihood, profile likelihood and confidence regions), the strategies used to build and sample the training set, the choice of training target, the SHAP interpretation of feature importance, and the Markov-Chain Monte Carlo exploration of the emulated posterior together with its closure tests. Section~\ref{sec:applications} presents three applications of the framework. We first analyse flavour anomalies in semileptonic $B$ meson decays, construct an XGBoost surrogate model of the \texttt{SMEFT19} global likelihood, and use it to map high-resolution confidence regions. We also rank the model parameters with SHAP, and study the correlations among low-energy observables that characterise the considered scenario. Then, we consider light NP, where the lightness of the new states induces non-Gaussian and even discontinuous likelihoods. As a concrete example, we explore axion-like particles as a candidate explanation for the Belle~II $B^+\to K^+\nu\bar\nu$ excess, training the surrogate through an active-learning programme and a two-stage classifier-plus-regressor strategy to resolve the relevant region of parameter space. Finally, we address cosmology, fitting the $\Lambda$CDM and dynamical dark-energy ($w_0w_a$CDM) models to Type~Ia supernovae, baryon acoustic oscillations, and a compressed cosmic-microwave-background prior. Because the exact cosmological likelihood is cheap enough to sample directly, this application doubles as a controlled validation of the method, in which the emulated posteriors are shown to be statistically indistinguishable from the exact ones. Our conclusions and perspectives are presented in Section~\ref{sec:conclusions}.

\section{Machine Learning Algorithm}\label{sec:ml}

ML algorithms are powerful tools for global fits both in high energy physics and cosmology. One central feature of the framework is the application of SHAP values to quantify the relative importance of each parameter in constraining the fit. This provides a model-independent ranking of model features based on their impact on specific Wilson coefficients of an effective field theory, and allows a transparent mapping between experimental inputs and the parameters of the effective field theories, going beyond standard $\chi^2$-based sensitivity analyses~\cite{Alda:2021rgt,Alda:2020okk,AldaGallo:2021cku,Penaranda:2024ssy,MirRamos:2025ijy,Alda:2026jnc}. With this type of analysis, we can identify the regions of parameter space allowed by any model compatible with current experimental data.

We adopt a ML approach to approximate the log-likelihood function. Specifically, we use the XGBoost algorithm~\cite{Chen:2016btl}, an ensemble method based on regression trees that can efficiently approximate arbitrary functions. Regression trees divide the parameter space into subsets and assign a numerical value to each leaf. By combining many trees into an ensemble, the model constructs an approximation of the log-likelihood function using supervised learning. The training dataset consists of parameter points and their corresponding precomputed log-likelihood values. Model optimization relies on a loss function that measures prediction accuracy and a regularization term that limits model complexity to avoid overfitting. The ensemble is built iteratively by adding new tree structures while applying shrinkage techniques to improve generalization. Using the machine-learned approximation instead of the exact likelihood significantly reduces computational costs, since Monte Carlo methods require a large number of likelihood evaluations.

The use of machine learning to accelerate or replace expensive likelihood evaluations has a rich and rapidly growing history across particle physics and cosmology. A large body of work is built on neural networks, used either as direct emulators of the likelihood or of the underlying observables, or within the broader programme of simulation-based (likelihood-free) inference, in which neural density estimators learn the likelihood or the posterior directly from forward simulations~\cite{Cranmer:2019eaq,Alsing:2019xrx,SpurioMancini:2021ppk}. These methods are powerful and highly flexible, but as universal function approximators, neural networks typically require large training sets and careful regularisation to avoid over-smoothing the narrow ridges and sharp transitions that characterise the non-Gaussian likelihoods considered here. We instead adopt gradient-boosted regression trees (XGBoost), which reach the required accuracy with comparatively modest training sets, remain stable under reseeding without extensive tuning, and admit an exact and efficient feature-attribution analysis through SHAP values, keeping the emulated inference interpretable. This choice, and the explicit comparison with neural-network surrogates, follows our earlier flavour analysis~\cite{Penaranda:2024ssy}.

SHAP values~\cite{Lundberg:2017uca,Lundberg:2018} are used to interpret the influence of individual parameters in the approximation. Based on concepts from cooperative game theory~\cite{Shapley:1953}, SHAP values quantify the contribution of each feature to the model prediction while satisfying properties such as local accuracy, missingness, and consistency. SHAP analysis evaluates the marginal impact of including each parameter in the model and provides insight into the relative importance of the different fit parameters.

The statistical framework summarised in this section follows Ref.~\cite{Alda:2026jnc}.  The central statistical object in a global fit is the likelihood function $\mathcal{L}(\bm\theta)$, defined as the conditional probability of observing the experimental data given the model parameters $\bm{\theta}$,
\begin{equation}
    \mathcal{L}(\bm\theta) = p(\text{data}\,|\,\bm{\theta})\,.
\end{equation}

It is convenient to work with $\chi^2(\bm{\theta}) \equiv -2\log\mathcal{L}(\bm{\theta})$.
For $N$ statistically independent observables with Gaussian uncertainties,
\begin{equation}
    \mathcal{L}(\bm{\theta}) = \prod_{i=1}^{N} \frac{1}{\sqrt{2\pi}\,\sigma_i}
    \exp\!\left(-\frac{(y_i - x_i(\bm{\theta}))^2}{2\,\sigma_i^2}\right)\,,
    \label{eq:likelihood}
\end{equation}
where $y_i$ are the experimental measurements, $x_i(\bm{\theta})$ are the theoretical predictions, and $\sigma_i^2 = \sigma_{y_i}^2 + \sigma_{x_i}^2$ is the sum of experimental and theoretical variances.  When the observables are correlated, the product is replaced by the multivariate Gaussian $\chi^2(\bm{\theta}) = \sum_{ij}(y_i - x_i(\bm{\theta}))\,(\mathcal{C}^{-1})_{ij}\,(y_j - x_j(\bm{\theta}))$, with $\mathcal{C}$ the combined experimental and theoretical covariance matrix. The best-fit point is the Maximum Likelihood Estimator (MLE),
\begin{equation}
    \hat{\bm{\theta}} = \arg\max_{\bm{\theta}}\,\mathcal{L}(\bm{\theta}) = \arg\min_{\bm{\theta}}\,\chi^2(\bm{\theta})\,,
\end{equation}
and the goodness of a new-physics hypothesis is quantified by its pull with respect to the Standard Model (SM), $\Delta\chi^2_{\rm SM} = \chi^2(\bm{\theta}_{\rm SM}) - \chi^2(\hat{\bm{\theta}})$.

Confidence regions are obtained via the profile likelihood. The profile likelihood for the $i$-th parameter is defined as
\begin{equation}
    \mathcal{L}_P(\theta^i) = \max_{\theta^{j \neq i}}\,\mathcal{L}(\theta^1,\ldots,\theta^n)\,,
\end{equation}
and the statistic test
\begin{equation}
    \lambda_P(\theta^i) = -2\log\frac{\mathcal{L}_P(\theta^i)}{\mathcal{L}(\hat{\bm{\theta}})}
    = \Delta\chi^2(\theta^i)
\end{equation}
follows asymptotically a $\chi^2$ distribution with one degree of freedom (Wilks' theorem~\cite{Wilks:1938dza}). The $k\,\sigma$ confidence interval on $\theta^i$ is therefore the set of values for which $\lambda_P(\theta^i) \leq k^2$. For two-dimensional confidence regions one uses the profile likelihood over the remaining parameters, and the $1\,\sigma$ and $2\,\sigma$ contours correspond to $\Delta\chi^2 = 2.30$ and $6.18$ respectively.

In practice, however, the likelihood in Eq.~(\ref{eq:likelihood}) is often only a simplified picture. Some applications considered in this work involve likelihoods that are non-Gaussian, depend on a large number of correlated observables, and require expensive theoretical computations for each prediction $x_i(\bm{\theta})$. Mapping the full $\chi^2$ surface therefore requires a large number of evaluations, making a brute-force scan computationally prohibitive. We address this by training a ML surrogate to approximate the $\chi^2$ function~\cite{Alda:2026jnc}.

The training dataset $\mathcal{D} = \{(\theta_i,\, y_i)\}_{i=1}^{N}$ is built by evaluating the exact $\chi^2$ at a set of parameter-space points. To distribute these points efficiently we combine a space-filling design over the full prior volume with a denser draw concentrated around the best-fit region, where the likelihood is largest and the confidence contours are most sensitive. For the space-filling component we use, depending on the application, plain uniform random sampling, Latin Hypercube Sampling (LHS)~\cite{McKay:1979}, or a low-discrepancy Sobol sequence~\cite{Sobol:1967}. The latter two guarantee substantially more uniform coverage than a purely random sample of the same size, while uniform sampling is adequate whenever a large point budget and the dense best-fit cloud already cover the relevant region. The denser best-fit draw is most efficiently realised as a Gaussian cloud whose covariance is derived from the Hessian of the fit, so that the sampling follows the actual likelihood ellipsoid and automatically resolves narrow, anisotropic degeneracies without a hand-tuned grid.  When the relevant region is not known a priori, or when the likelihood develops sharp or narrow structures that a static design samples too coarsely, this base design is refined iteratively through active learning.  At each iteration an inexpensive auxiliary model is trained on the data accumulated so far, typically a Gaussian Process (GP), the infinite-dimensional limit of a multivariate Gaussian, which returns both a prediction and an uncertainty at every point. The next points to be evaluated exactly are then chosen by maximising an acquisition function such as the Expected Improvement, which balances exploration of the regions where the surrogate is most uncertain against exploitation of the regions of lowest $\chi^2$.  The exact $\chi^2$ is computed at the selected points, they are appended to the training set, and the cycle repeats, progressively concentrating the sampling effort where it most improves the emulated likelihood.  

The surrogate model is an XGBoost gradient-boosted ensemble of regression trees~\cite{Chen:2016btl}. A single regression tree partitions the parameter space into $T$ disjoint regions (``leaves'') through a sequence of binary cuts and assigns a constant weight to each one, defining a function $f(\theta) = w_{q(\theta)}$, where $q(\theta)\in\{1,\ldots,T\}$ is the leaf into which $\theta$ falls and $w_{q(\theta)}$ its weight. Since a single tree is too rigid to reproduce an arbitrary function, the surrogate is built as an ensemble of $K$ trees,
\begin{equation}
    \phi(\bm{\theta}) = \sum_{k=1}^{K} f^{(k)}(\bm{\theta}) = \sum_{k=1}^{K} w^{(k)}_{q(\bm{\theta})}\,,
    \label{eq:ensemble}
\end{equation}
whose output $\phi(\bm{\theta})$ is the approximation of the training target (the $\chi^2$ or the log-likelihood). The trees are obtained by supervised learning on the dataset $\mathcal{D}$, by minimising the objective
\begin{equation}
    \mathrm{Obj}[\phi] = \sum_i \ell\big(\phi(\theta_i),\,y_i\big) + \sum_k \Omega\big(f^{(k)}\big)\,,
    \label{eq:xgb_objective}
\end{equation}
which balances a loss function against a regularisation term. The loss $\ell$ measures the discrepancy between the target $y_i$ and its approximation $\phi(\theta_i)$; we use the mean absolute error, $\ell(\phi(\theta_i),\,y_i) = |\phi(\theta_i) - y_i|$, which is more robust to the large-$\chi^2$ outliers than a quadratic loss. The regularisation $\Omega$ penalises the complexity of each tree,
\begin{equation}
    \Omega(f) = \gamma\,T + \tfrac{1}{2}\lambda\,\|w\|^2\,,
    \label{eq:xgb_reg}
\end{equation}
with $T$ the number of leaves and $w$ the vector of leaf weights, so that the hyperparameters $\gamma$ and $\lambda$ introduced below directly control the trade-off between flexibility and overfitting. XGBoost additionally provides an optional $L_1$ penalty $\alpha\,\|w\|_1$ on the leaf weights (the hyperparameter $\alpha$ below), which we employ in the axion-like-particle fit. The ensemble is built iteratively: starting from a single tree that captures the overall shape of the $\chi^2$ surface, each subsequent tree provides an incremental correction whose newly added weights are scaled (shrinkage) by a learning rate $\eta < 1$. Training is stopped early as soon as the validation loss stagnates for a fixed number of consecutive rounds. Because regression trees partition the parameter space adaptively rather than assuming global smoothness, they are particularly well suited to likelihoods with narrow ridges, sharp transitions and plateau-like regions, where neural-network surrogates would require significantly larger training sets and careful regularisation to avoid over-smoothing the fine structure.

The behavior of the model is controlled by a set of hyperparameters:
\begin{itemize}
    \item \textbf{Number of estimators}: number of trees in the ensemble.
    \item \textbf{Maximum depth}: maximum number of levels per tree.  Large
          values increase expressiveness but also the tendency to overfit.
    \item \textbf{Learning rate} ($\eta$): scales the contribution of each
          tree.  Smaller values require more trees but reduce overfitting.
    \item \textbf{Minimum split loss} ($\gamma$): minimum loss reduction
          required to create a new partition in a leaf.  Larger values
          yield more conservative trees.
    \item \textbf{Regularisation parameters} ($\alpha$, $\lambda$): penalise
          large leaf weights, further preventing overfitting.
    \item \textbf{Subsampling fractions} (\texttt{subsample},
          \texttt{colsample\_bytree}): proportion of training rows and
          feature columns used per tree, introducing randomness and
          reducing variance.
\end{itemize}

These hyperparameters can be tuned automatically using Optuna~\cite{Akiba:2019}, a gradient-free optimisation library based on a tree-structured Parzen estimator, although for low-dimensional, smooth landscapes a fixed, conservative configuration (moderate depth, small learning rate, aggressive early stopping) is often more robust. The choice of training target (direct $\chi^2$, sigmoid transform, or shifted-log$_{10}$ transform) depends on the structure of the likelihood landscape and is discussed in each application section.  Because tree ensembles do not extrapolate reliably beyond their training range, the surrogate is trusted only within the sampled domain: the space-filling component of the training design covers the full prior volume, and the sampling density is increased wherever the landscape is most structured, so that the emulated likelihood remains faithful across the whole parameter space.

After training, SHAP values~\cite{Lundberg:2017uca,Lundberg:2018} are computed to quantify the individual contribution of each parameter to the model output at every point in the parameter space. Denoting by $F$ the full set of $n$ features (the fit parameters), the SHAP value of the $j$-th feature at a point $\bm{\theta}$ is its average marginal contribution over all subsets $S$ of the remaining features~\cite{Lundberg:2017uca,Lundberg:2018},
\begin{equation}
    \phi_j(\bm{\theta}) = \sum_{S\subseteq F\setminus\{j\}} \frac{|S|!\,(|F|-|S|-1)!}{|F|!}\,\big[\,f_{\bm{\theta}}(S\cup\{j\}) - f_{\bm{\theta}}(S)\,\big]\,,
    \label{eq:shap}
\end{equation}
where $f_{\bm{\theta}}(S)$ is the prediction obtained using only the features in the subset $S$. These attributions satisfy three key properties: local accuracy, missingness, and consistency. Local accuracy means that they decompose the model output exactly,
\begin{equation}
    \phi(\bm{\theta}) = \phi_0 + \sum_{j=1}^{n} \phi_j(\bm{\theta})\,,
    \label{eq:shap_localacc}
\end{equation}
with $\phi_0$ the base value, equal to the average prediction over the dataset. Missingness dictates that a feature missing from the input space or irrelevant is assigned a null contribution ($\phi_j = 0$). Consistency guarantees that if the model changes such that the marginal contribution of a feature increases or remains the same regardless of the other features, its corresponding SHAP value will not decrease.

Evaluating Eq.~(\ref{eq:shap}) directly would require retraining the model on every feature subset, an operation whose computational cost grows exponentially with the number of features. For tree ensembles, however, the exact SHAP values are recovered in polynomial time by the TreeSHAP algorithm~\cite{Lundberg:2018}. Aggregated over the training dataset, SHAP values provide a global, model-independent ranking of parameter importance that complements the standard $\Delta\chi^2$ sensitivity analysis and reveals non-linear dependencies and correlations between parameters.

To map the posterior distribution, the emulated $\chi^2$ is sampled with a Markov-Chain Monte Carlo (MCMC) algorithm. A tentative point $\bm{\theta}'$ is accepted over the current one with the Metropolis criterion
\begin{equation}
    \log u < \log\mathcal{L}(\bm{\theta}') - \log\mathcal{L}(\bm{\theta})\,,\qquad u\sim\mathcal{U}(0,1)\,,
    \label{eq:metropolis}
\end{equation}
with $u$ a random number uniformly distributed between $0$ and $1$; the proposal uses a Gaussian kernel whose covariance is, after a burn-in phase, set to the Cholesky factor of the fit covariance, substantially improving mixing in correlated parameter spaces. A parallel Random-Walk Metropolis--Hastings variant running $n_{\rm chains}$ independent chains can be used, with sampling stopped automatically once the effective sample size exceeds a predefined target. A useful closure test is the distribution of $\Delta\chi^2 = 2\,[\log\mathcal{L}(\hat{\bm{\theta}}) - \log\mathcal{L}(\bm{\theta})]$ over the accepted points: in the Gaussian limit, where the log-likelihood is locally quadratic, it follows a $\chi^2$ distribution with a number of degrees of freedom equal to the number of fit parameters (Wilks' theorem~\cite{Wilks:1938dza}), so that any departure from this reference is itself a quantitative diagnostic of the non-Gaussianity of the likelihood.  Because each forward pass through the trained ensemble replaces a full physics-library evaluation, the surrogate reduces the wall-clock time of the posterior exploration by several orders of magnitude.

\section{Applications}\label{sec:applications}

\subsection{\texorpdfstring{$B$ Physics}{B Physics}}\label{sec:bphysics}

Flavour physics is an exceptionally sensitive testing ground for the Standard Model. Because flavour-changing neutral currents (FCNCs) are highly suppressed within the SM framework, they are uniquely susceptible to the virtual contributions of heavy particles. In recent years, a series of persistent discrepancies between SM predictions and experimental data, known as flavour anomalies, have intensified interest in this sector~\cite{LHCb:2017vlu,LHCb:2021trn,HFLAV:CKM2025}. To analyze these anomalies without committing to a single specific ultraviolet theory, effective field theories are a very useful tool~\cite{Alda:2020okk,Alda:2021rgt,AldaGallo:2021cku}. By systematically integrating out heavy degrees of freedom, such as hypothetical NP mediators, we are left with an effective theory of contact interactions, governed by operators and their associated coupling  strengths, known as Wilson Coefficients.

One of the most powerful features of flavour physics is that a single NP mediator cannot alter an isolated observable without triggering a cascade of predictable shifts in others. Global fits exploit these correlations~\cite{Alda:2021rgt}.

Semileptonic $B$ meson decays provide some of the most stringent low-energy tests of the SM.  Over the last decade a coherent set of deviations, the flavour anomalies, has accumulated in three classes of transitions: the charged-current $b\to c\tau\nu$ ratios $R_{D}$, $R_{D^*}$ and $R_{J/\psi}$; the neutral-current $b\to s\nu\bar\nu$ branching ratios, most notably the $2.8\,\sigma$ excess in $\mathrm{BR}(B^+\!\to K^+\nu\bar\nu)$ reported by Belle~II~\cite{Belle-II:2023esi}; and the $b\to s\ell^+\ell^-$ lepton-flavour-universality ratios $R_{K}$, $R_{K^*}$. Following our previous global analyses~\cite{Alda:2020okk,Alda:2021rgt,AldaGallo:2021cku,Penaranda:2024ssy}, the short-distance NP is parameterised through the Wilson coefficients of the SMEFT, defined at the scale $\Lambda = 1\,\mathrm{TeV}$ in the Warsaw basis and evolved down to the scale of each observable, and the fit is performed with the \texttt{SMEFT19} package~\cite{SMEFT19}, which builds the global likelihood from more than a hundred observables by interfacing \texttt{flavio}~\cite{Straub:2018kue}, \texttt{smelli}~\cite{Aebischer:2018iyb} and \texttt{wilson}~\cite{Aebischer:2018bkb}. Within this pipeline the three packages play complementary roles: \texttt{wilson} performs the renormalisation-group evolution of the Wilson coefficients from the SMEFT scale down to the low-energy (WET) scale and the change between operator bases, \texttt{flavio} computes the theoretical predictions of the low-energy observables from the resulting coefficients, and \texttt{smelli} assembles the global likelihood by comparing these predictions with the experimental measurements and their combined experimental and theoretical covariance.

\subsubsection{Scenario~III of the SMEFT Global Fit}

We work in Scenario~III of the mass-rotation framework introduced in Ref.~\cite{Penaranda:2024ssy}, in which the singlet and triplet four-fermion operators carry independent Wilson coefficients $C_1\neq C_3$ and no mixing is allowed in the lepton sector ($\alpha^\ell=\beta^\ell=0$). Explicitly, the new-physics interactions are described by the dimension-six effective Lagrangian~\cite{Penaranda:2024ssy},
\begin{equation}
    \mathcal{L}_{\rm NP} = \frac{\lambda^\ell_{ij}\,\lambda^q_{kl}}{\Lambda^2}\Big[\, C_1\,\big(\bar\ell_i \gamma_\mu \ell_j\big)\big(\bar q_k \gamma^\mu q_l\big) + C_3\,\big(\bar\ell_i \gamma_\mu \tau^I \ell_j\big)\big(\bar q_k \gamma^\mu \tau^I q_l\big)\,\Big]\,,
    \label{eq:smeft_scIII}
\end{equation}
where $\ell_i$ and $q_i$ are the $SU(2)_L$ doublets for the $i$-th generation of leptons and quarks, respectively, $\Lambda = 1\,\mathrm{TeV}$ is the new-physics scale, the two structures are the singlet $\mathcal{O}^{(1)}_{\ell q}$ and triplet $\mathcal{O}^{(3)}_{\ell q}$ four-fermion operators of the Warsaw basis, and $\lambda^\ell$ and $\lambda^q$ are the lepton- and quark-flavour mixing matrices. Each of them encodes the assumption that the new physics couples to a single (third-generation) direction in the interaction basis and is then rotated to the mass basis; it is the idempotent, Hermitian rank-one matrix
\begin{equation}
    \lambda^f = \frac{1}{1+|\alpha^f|^2+|\beta^f|^2}
    \begin{pmatrix}
        |\alpha^f|^2 & \alpha^f\beta^{f*} & \alpha^f\\
        \alpha^{f*}\beta^f & |\beta^f|^2 & \beta^f\\
        \alpha^{f*} & \beta^{f*} & 1
    \end{pmatrix}\,,\qquad f=\ell,q\,,
    \label{eq:idemp}
\end{equation}
built from the direction $(\alpha^f,\beta^f,1)$, so that $\alpha^f=\beta^f=0$ recovers a pure third-generation coupling. Scenario~III fixes $\alpha^\ell=\beta^\ell=0$ (no lepton mixing, $\lambda^\ell=\mathrm{diag}(0,0,1)$) and $\alpha^q=0$, so that mixing occurs only between the second and third quark generations.  The fit therefore depends on three parameters $(C_1,\,C_3,\,\beta^q)$, with $\beta^q$ being the quark mixing angle between the second and third generations.  The relevant transitions probe different combinations of the coefficients: $b\to s\nu\bar\nu$ receives tree-level contributions proportional to $C_1-C_3$, $b\to c\ell\nu$ is governed by $C_3$, and $b\to s\ell^+\ell^-$ depends on $C_1+C_3$.  Treating $C_1$ and $C_3$ as independent, unlike Scenarios~I and~II where $C_1=C_3$, is precisely what allows the fit to accommodate the enhanced $B^+\!\to K^+\nu\bar\nu$ rate together with the $R_{D^{(*)}}$ anomalies. The parameters are scanned over $C_1,\,C_3\in[-0.30,\,0.00]$ and $\beta^q\in[0.00,\,3.20]$.

With respect to Ref.~\cite{Penaranda:2024ssy} we update the dominant charged- and neutral-current inputs to their most recent determinations. For the $b\to c\tau\nu$ ratios we use the HFLAV CKM~2025 world averages~\cite{HFLAV:CKM2025}, $R_D = 0.358\pm0.024$ and $R_{D^*} = 0.281\pm0.011$ with a correlation coefficient of $-0.37$, supplemented by the two $R_{J/\psi}$ measurements, $0.71\pm0.17\pm0.18$ from LHCb~\cite{LHCb:2017vlu} and $0.49\pm0.26$ from CMS~\cite{CMS:2025jfx}, which we include as independent constraints. For $b\to s\nu\bar\nu$ we include the Belle~II $362\,\mathrm{fb}^{-1}$ result $\mathrm{BR}(B^+\!\to K^+\nu\bar\nu) = (2.3\pm0.7)\times10^{-5}$~\cite{Belle-II:2023esi}.  The $b\to s\ell^+\ell^-$ ratios $R_K$, $R_{K^*}$, now in agreement with the SM, are kept at their reference values and do not move the fit, since lepton-flavour universality is exact in Scenario~III~\cite{Penaranda:2024ssy}.

\subsubsection{Surrogate Training and Validation}

\begin{figure}[H]
    \centering
    \includegraphics[width=0.42\linewidth]{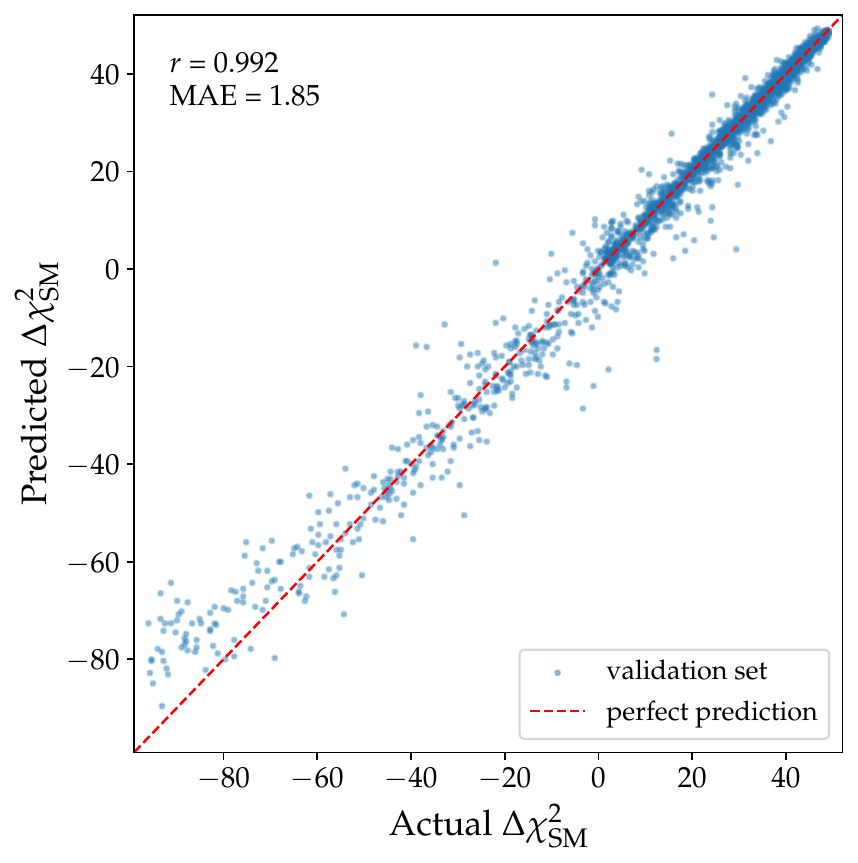}\\
    \makebox[0.42\linewidth]{(a)}\\[1.5ex]
    \includegraphics[width=0.49\linewidth]{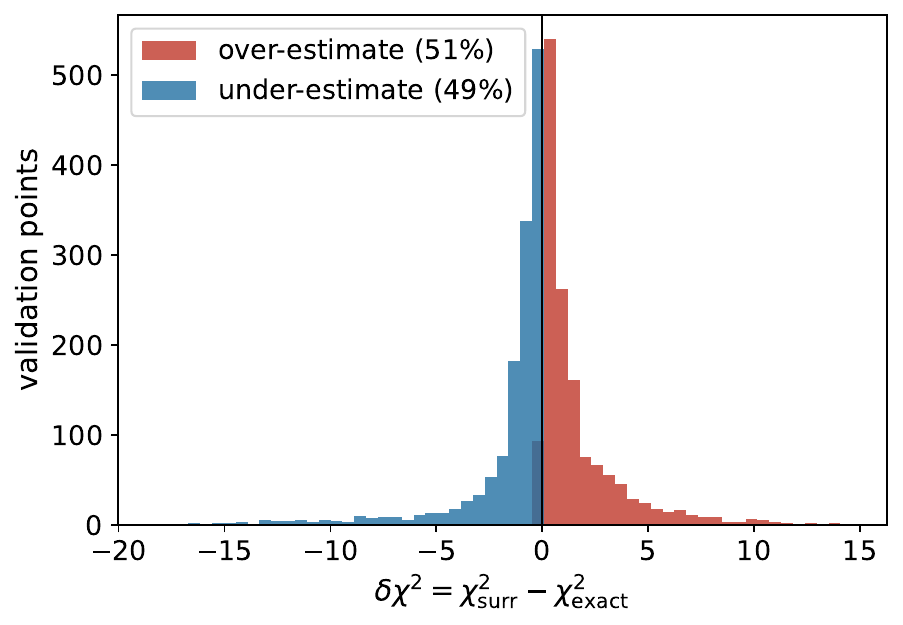}\hfill
    \includegraphics[width=0.50\linewidth]{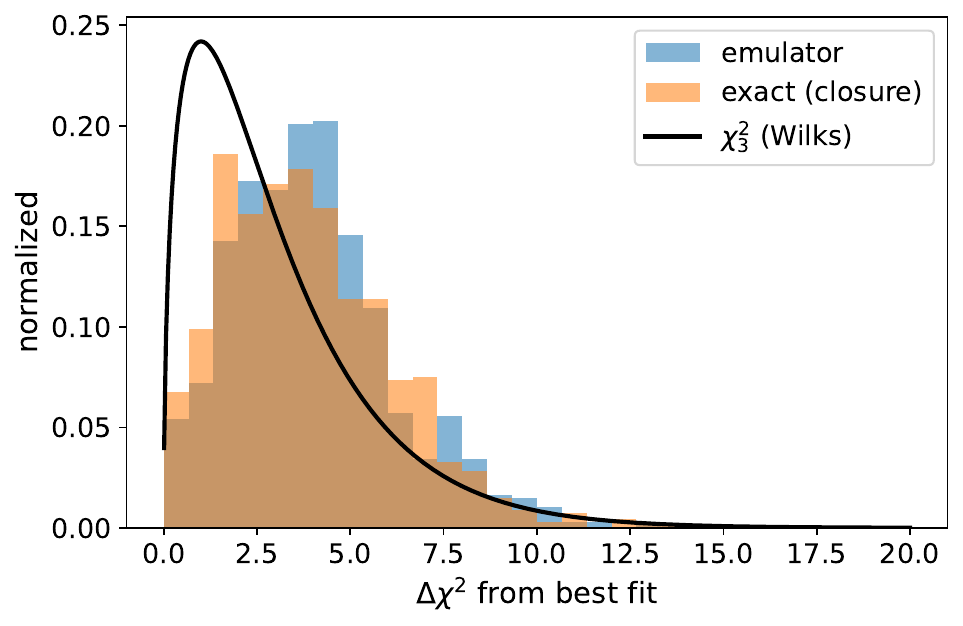}\\
    \makebox[0.49\linewidth]{(b)}\hfill\makebox[0.50\linewidth]{(c)}
    \caption{Validation of the $B$-physics emulator in Scenario~III.
    {(\textbf{a})} Predicted versus exact $\Delta\chi^2_{\rm SM}$ on the held-out validation set.
    {(\textbf{b})} Signed residuals $\delta\chi^2=\chi^2_{\rm surr}-\chi^2_{\rm exact}$ on the validation set, split into over- (red) and under-estimates (blue).
    {(\textbf{c})} Posterior-level closure test: distribution of $\Delta\chi^2=\chi^2-\chi^2_{\rm bf}$ over a sample drawn from the emulated posterior, scored with the emulator (blue) and by re-evaluating the exact \texttt{SMEFT19} likelihood at the same points (orange), together with the $\chi^2$ distribution for three degrees of freedom (Gaussian/Wilks limit, black).}
    \label{fig:bphys_perf}
\end{figure}

We train an XGBoost surrogate of the \texttt{SMEFT19} global log-likelihood over the three-dimensional space $(C_1,C_3,\beta^q)$, following the methodology of Section~\ref{sec:ml} and first applied to this problem in Refs.~\cite{Alda:2021rgt,Penaranda:2024ssy}. The training set comprises a base design of $\sim\!1.4\times10^{4}$ points: a Sobol space-filling sample over the full domain, a Gaussian cloud drawn with the fit covariance around the best-fit point (which resolves the narrow $C_1$ direction automatically), and three two-dimensional grids used to validate the likelihood maps. This base design is supplemented by a further $\sim\!1.2\times10^{4}$ exact evaluations concentrated, through an active-learning refinement, in the weakly constrained high-$\beta^q$ region where it is sparse, bringing the total to $\sim\!2.6\times10^{4}$ points. Densifying this region improves the coverage of the extended $\beta^q$ ridge and mitigates the over-extrapolation of the surrogate where the training data would otherwise be scarce. Trained directly on the log-likelihood with a small learning rate and aggressive early stopping, the surrogate reaches a Pearson correlation coefficient $r = 0.99$ ($R^2 = 0.98$) on a held-out validation set (Figure~\ref{fig:bphys_perf}a) while evaluating each point far faster than the exact likelihood, as quantified below.

\begin{figure}[H]
    \centering
    \includegraphics[width=\linewidth]{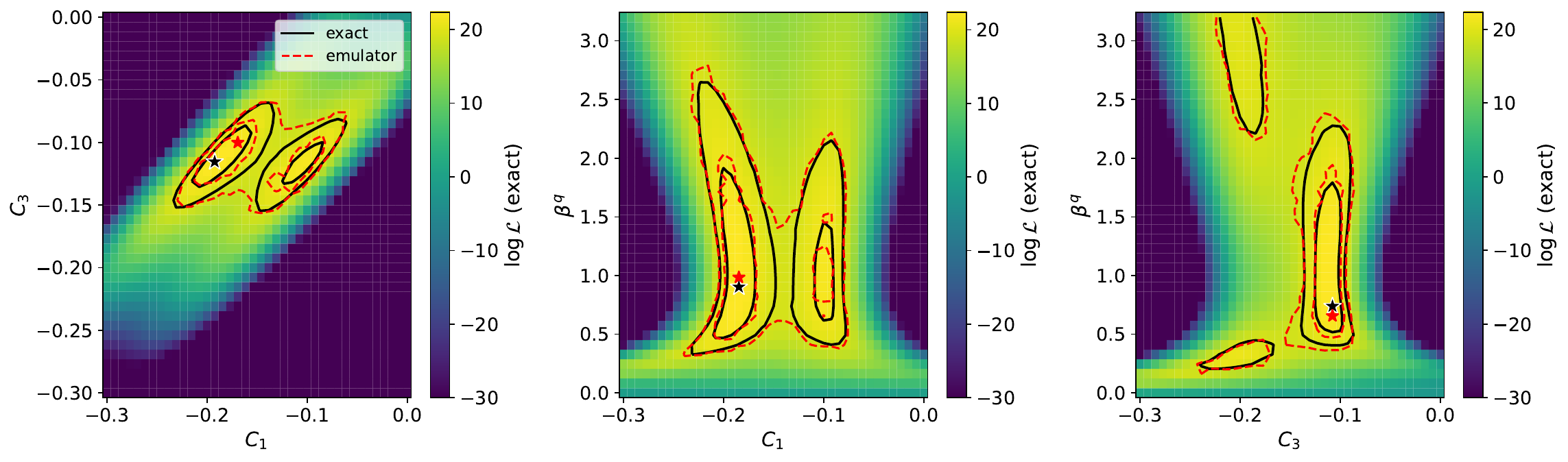}
    \caption{Exact \texttt{SMEFT19} log-likelihood maps in the $(C_1,C_3)$, $(C_1,\beta^q)$ and $(C_3,\beta^q)$ planes, each with the third parameter fixed at the best fit, on a $40\times40$ grid.  Overlaid are the $1\,\sigma$ and $2\,\sigma$ profile contours ($\Delta\chi^2=2.30,\,6.18$) of the exact likelihood (black, solid) and of the XGBoost emulator (red, dashed); the stars mark the exact (black) and emulator (red) best fits.  The emulated contours reproduce the exact ones, including the extended, curved $\beta^q$ ridge.}
    \label{fig:bphys_maps}
\end{figure}

The decisive advantage of the surrogate is its evaluation speed, quantified in Table~\ref{tab:bphys_timing}.  All times quoted here are measured on a single CPU core, so the comparison is reproducible and does not depend on the number of cores or on the specific machine.  Once trained, the XGBoost ensemble predicts the log-likelihood in $\sim\!34\,\upmu$s per point, to be compared with $\sim\!3$\,s for a single evaluation of the exact \texttt{SMEFT19} global likelihood, which requires running the SMEFT-to-WET renormalisation-group evolution and recomputing more than a hundred observables at each point. This is a speed-up of about $1.0\times10^{5}$ per evaluation.  The gain is most tangible in the construction of the training set: generating the $\sim\!2.6\times10^{4}$ exact points that define the emulator costs about $24$ core-hours of \texttt{SMEFT19} computation, whereas the trained surrogate re-evaluates the same number of points in $\sim\!0.9$\,s.  It is precisely this gain that makes the high-resolution confidence contours and the SHAP sensitivity analysis computationally feasible, while the agreement of the emulated likelihood maps with the exact ones (Figure~\ref{fig:bphys_maps}) guarantees that the speed-up does not come at the price of accuracy.

\begin{table}[H]

\caption{Evaluation cost of the XGBoost emulator against the exact \texttt{SMEFT19} global likelihood in Scenario~III, all on a single CPU core.  Per-evaluation times use $10^{4}$ surrogate predictions (batch) and a serial run of distinct exact evaluations; the second row is the cost of generating the full \mbox{$\sim\!2.6\times10^{4}$-point} training set.  Single-core figures make the speed-up reproducible and independent of the core~count.}
\label{tab:bphys_timing}
 
\setlength{\tabcolsep}{5.3mm}
\resizebox{\textwidth}{!}{\begin{tabular}{lccc}
\toprule
 & \textbf{XGBoost Emulator} & \textbf{Exact {\texttt{SMEFT19}}} & \textbf{Speed-Up }\\
\midrule
Per evaluation  & $34\,\upmu$s  & $3.3$\,s     & $1.0\times10^{5}$ \\
Training set ($2.6\times10^{4}$ points) & $0.88$\,s   & $24$\,core-h & $1.0\times10^{5}$ \\
\bottomrule
\end{tabular}}

\end{table}

\vspace{-12pt}

The fit locates the best-fit point at
\begin{equation}
    C_1 = -0.185\,,\qquad C_3 = -0.110\,,\qquad \beta^q = 0.83\,,
\end{equation}
with $\Delta\chi^2_{\rm SM} = 48.8$, corresponding to a pull above $6\,\sigma$ with respect to the SM in the asymptotic (Wilks) limit. This significance should be read as nominal rather than exact: the $\chi^2$ calibration of the pull assumes a Gaussian likelihood, whereas the present one is markedly non-Gaussian (as the $\Delta\chi^2$ distribution of Figure~\ref{fig:bphys_perf}c below makes explicit), and the SM point $C_1=C_3=0$ lies on the boundary of the scanned range $C_1,C_3\in[-0.30,0]$, where the asymptotic distribution of the test statistic is modified into a mixture of $\chi^2$ distributions and the naive $\chi^2$ $p$-value over-estimates the significance. We therefore quote the $6\,\sigma$ as a nominal, asymptotic significance; establishing its exact value would require a dedicated calibration of the null distribution with SM pseudo-experiments, which lies beyond the scope of this methodological study and does not affect the qualitative preference for Scenario~III over the SM.  The fit clearly breaks the $C_1=C_3$ degeneracy, confirming the qualitative picture of Scenario~III with the updated data. We validate the surrogate specifically in the statistically relevant low-$\Delta\chi^2$ region.  On the held-out validation set the signed residuals $\delta\chi^2=\chi^2_{\rm surr}-\chi^2_{\rm exact}$ are almost perfectly balanced, $51\%$ over- and $49\%$ under-estimates with a negligible mean bias ($+0.04$ in $\chi^2$ units), so the emulator shows no systematic tendency to over- or under-cover (Figure~\ref{fig:bphys_perf}b).  Its global mean absolute error, $\mathrm{MAE}=1.85$, is dominated by the sparsely sampled high-$\Delta\chi^2$ tail and drops sharply in the band that sets the confidence contours, to $0.60$ within $2\,\sigma$ and $0.50$ within $1\,\sigma$ (Table~\ref{tab:bphys_validation}): the emulator is three to four times more accurate exactly where it matters.

\begin{table}[H]
\caption{Emulator error (in $\chi^2$ units) on the held-out validation set, globally and restricted to the low-$\Delta\chi^2$ region that defines the confidence contours.  The error in the statistically relevant band is three to four times smaller than the global figure.}
\label{tab:bphys_validation}

\setlength{\tabcolsep}{9.3mm}
\resizebox{\textwidth}{!}{\begin{tabular}{lccc}
\toprule
\textbf{Region (Exact \boldmath$\Delta\chi^2$ from Best Fit)} & \boldmath$N$ & \textbf{MAE} & \textbf{RMSE} \\
\midrule
all validation points                       & 2887 & 1.85 & 3.64 \\
within $2\,\sigma$ ($\Delta\chi^2\le6.18$)   & 533  & 0.60 & 0.85 \\
within $1\,\sigma$ ($\Delta\chi^2\le2.30$)   & 208  & 0.50 & 0.77 \\
\bottomrule
\end{tabular}}

\end{table}

The two-dimensional log-likelihood maps of Figure~\ref{fig:bphys_maps} confirm this at the level of the confidence regions.  The quark-mixing angle $\beta^q$ develops an extended, curved ridge in the $(C_1,\beta^q)$ and $(C_3,\beta^q)$ planes, far from the compact Gaussian ellipse that a quadratic log-likelihood would produce, and the $1\,\sigma$/$2\,\sigma$ contours of the emulator track the exact ones closely: the emulator $2\,\sigma$ area exceeds the exact one by only $19$--$29\%$ (a mild over-coverage, so the emulated contours are conservative rather than too narrow) and the best fit shifts by at most $8\%$ of the scanned range in each plane.

Finally, we perform a posterior-level closure test (Figure~\ref{fig:bphys_perf}c): we draw a representative sample from the emulated posterior and re-evaluate the exact \texttt{SMEFT19} likelihood at those same points.  The emulator and exact $\Delta\chi^2$ distributions overlap closely, with a root-mean-square difference of only $1.4$ in $\chi^2$ units within the $2\,\sigma$ band and $1.0$ within $1\,\sigma$, so the surrogate reproduces the exact likelihood over the region the posterior actually occupies.  Both distributions are displaced towards somewhat larger $\Delta\chi^2$ than the $\chi^2_3$ (Wilks) reference; because the exact and emulated histograms agree, this displacement is unambiguously a genuine imprint of the non-Gaussian likelihood (the weakly constrained $\beta^q$ ridge), not an emulator artefact.  The $\chi^2_3$ curve should therefore be read as the Gaussian reference recovered in the asymptotic limit, not as an expectation the distribution is required to match. Taken together, the signed residuals, the low-$\Delta\chi^2$ error budget (Table~\ref{tab:bphys_validation}), the contour comparison (Figure~\ref{fig:bphys_maps}) and this closure test establish that in the flavour application the surrogate is not merely an accelerator, which is what makes the high-resolution contours and the SHAP analysis computationally feasible, but also a statistically faithful replacement of the exact likelihood in the low-$\Delta\chi^2$ region that sets the confidence limits.

\subsubsection{Feature Importance and Observable Correlations}

\begin{figure}[t]
    \centering
    \includegraphics[width=0.7\linewidth]{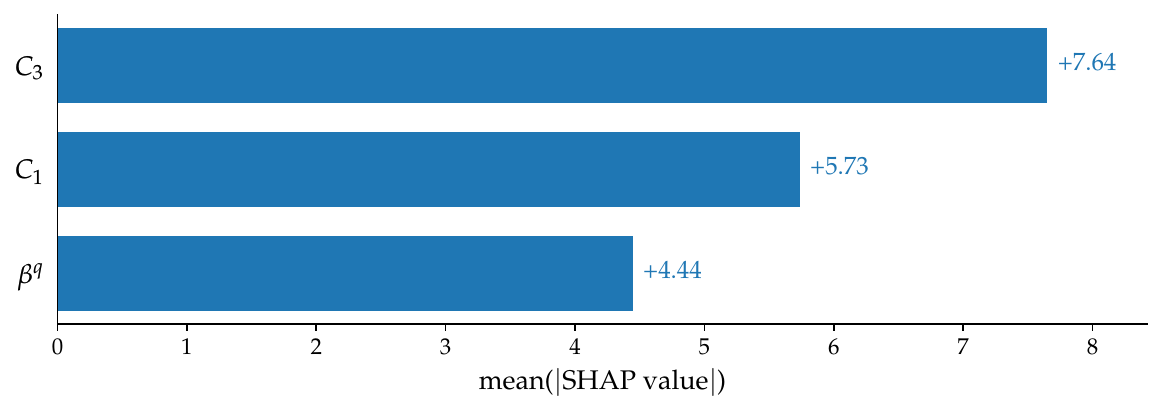}
    \caption{Global SHAP feature importance (mean absolute SHAP value) for the parameters of the $B$-physics fit in Scenario~III.}
    \label{fig:bphys_shap}
\end{figure}

The SHAP analysis (Figure~\ref{fig:bphys_shap}) ranks the parameters as $C_3 > C_1 > \beta^q$.  This hierarchy aligns with the underlying physics: the dominant sensitivity of the global fit comes from the charged-current $R_{D^{(*)}}$ observables (controlled by $C_3$), followed by the neutral-current $B\to K^{(*)}\nu\bar\nu$ modes (controlled by $C_1$), while the quark-mixing angle $\beta^q$ modulates both sectors without dominating. The agreement between the data-driven SHAP ranking and the expected operator structure confirms that the surrogate captures the correct EFT dynamics rather than overfitting numerical noise.

\begin{figure}[t]
    \centering
    \includegraphics[width=0.75\linewidth]{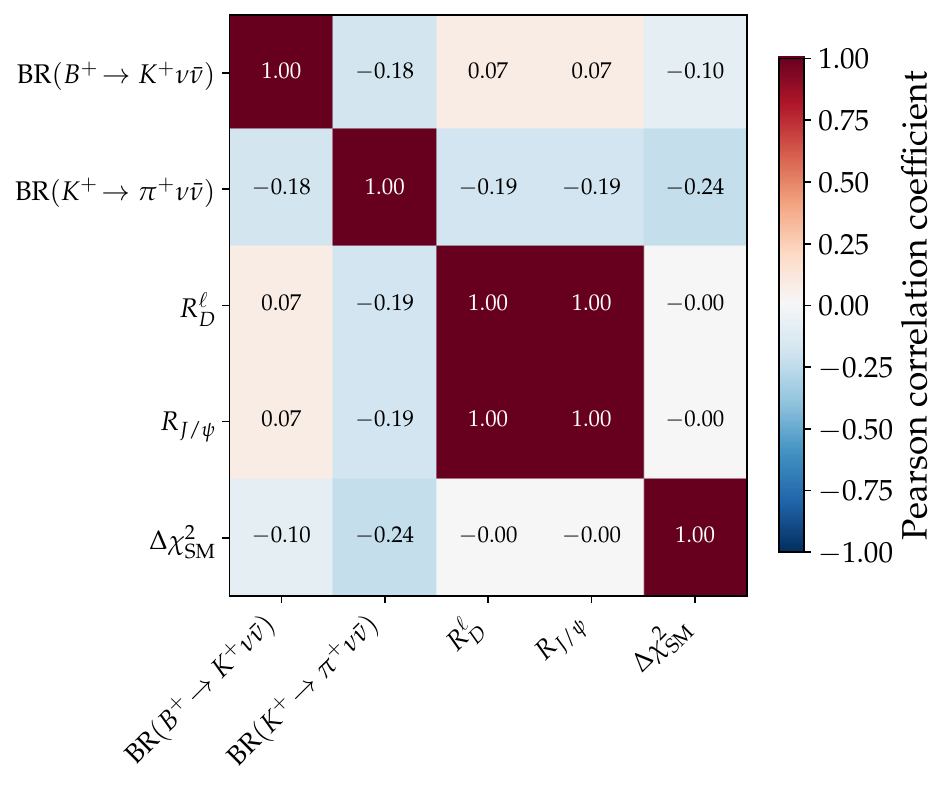}
    \caption{Matrix of Pearson correlation coefficients between the flavour observables that vary in Scenario~III, computed on the $1.5\times10^{4}$-point Monte Carlo sample generated with the emulator. Observables that remain SM-like in Scenario~III ($b\to s\ell^+\ell^-$ and $B_{(s)}\to\mu^+\mu^-$) carry no variation and are omitted.}
    \label{fig:bphys_corr}
\end{figure}

Using a sample of $1.5\times10^{4}$ points generated with the emulator, we run the SMEFT-to-WET renormalization-group evolution at each point and compute the Pearson correlation coefficients between the resulting low-energy observables (Figure~\ref{fig:bphys_corr}). Because no NP enters the lepton sector in Scenario~III ($\alpha^\ell=\beta^\ell=0$, so that $C_{10}^\mu=0$ and $R_{K^{(*)}}=1$ exactly), the $b\to s\ell^+\ell^-$ ratios and the $B_{(s)}\to\mu^+\mu^-$ branching ratios remain at their SM values and show no variation across the sample; the matrix therefore retains only the observables that genuinely respond to $(C_1,C_3,\beta^q)$.  The $b\to c\tau\nu$ ratios $R_D$ and $R_{J/\psi}$ are perfectly correlated, reflecting their common origin in the single coefficient $C_{V_L}\propto C_3$, while the neutral-current modes $\mathrm{BR}(B^+\!\to K^+\nu\bar\nu)$ and $\mathrm{BR}(K^+\!\to\pi^+\nu\bar\nu)$ respond to the orthogonal combination $C_1-C_3$.  The central result concerns the interplay between $R_D$ and $\mathrm{BR}(B^+\!\to K^+\nu\bar\nu)$: the almost-perfect correlation between these two observables found in Scenario~II~\cite{Alda:2021rgt} disappears (it drops to $\approx0.06$) once $C_1$ and $C_3$ are allowed to vary independently, since $\mathrm{BR}(B^+\!\to K^+\nu\bar\nu)\propto (C_1-C_3)$ while $R_D\propto C_3$.  The lack of a fixed correlation between the enhanced $B^+\!\to K^+\nu\bar\nu$ rate and the $b\to c\tau\nu$ ratios is a distinctive prediction of Scenario~III and a target for future experimental scrutiny.

\subsection{Light New Physics}
\label{sec:lightnp}

If the NP candidate is light and feebly interacting, there is a variety of new challenges, both in the experimental and in the theoretical arenas, arising from the fact that the new particles are physical states detectable in experiments (as opposed to the heavy case, where the new particles appear only as virtual particles in internal lines of Feynman diagrams). In general, we can distinguish several kinds of experimental signatures~\cite{Ferber:2022rsf,Bruggisser:2023npd,Alda:2025uwo}:
\begin{itemize}
    \item \textbf{Off-shell processes} where the corresponding on-shell process is forbidden by Lorentz invariance (e.g. leptonic and radiative meson decays $M \to X^* \to \ell^+\ell^-, \gamma\gamma$ and neutral meson mixing $M \to X^* \to \overline M$). These processes are able to impose bounds for all values of mass and lifetime of the new particle, although they are most relevant in the resonant region $m_X = m_M$.
    \item \textbf{Prompt decays:} the particle is produced in a reaction $A \to B X$ and subsequently decays into $X\to C$ (where $A$, $B$, $C$ are states that may contain multiple physical particles), and the experiment is not able to resolve the production and decay vertices. The experimental signature is therefore an excess in the rate of $A\to B C$ as compared to the SM prediction. If the experiment is able to resolve a distance $L_\mathrm{min}$, then the probability for the new particle to travel less than $L_\mathrm{min}$ and decay promptly is
    \begin{equation}
        P_\mathrm{prompt} = 1 - \exp\left(\frac{-L_\mathrm{min}}{c\tau_X \beta_X \gamma_X}\right)\,,
    \end{equation}
    where $c\tau_X = c\hbar/\Gamma_X$ is the proper decay length of the particle, and $\beta_X$ and $\gamma_X$ its velocity and Lorentz boost in the lab frame. If the Narrow Width Approximation (NWA) is valid $\Gamma_X \ll m_X$, then the NP prediction for this process is given by
    \begin{equation}
        \sigma(A\to BC)_\mathrm{NWA} = \sigma(A\to BX)\times \mathrm{BR}(X\to C) \times P_\mathrm{prompt}\,.
    \end{equation}
    \item \textbf{Displaced vertex decays:} the particle is produced in $A\to BX$ and decays as $X\to C$, but the experiment is able to resolve the production and decay vertices. This is the case of experiments with excellent vertex resolution, like LHCb, Belle II and BESIII. The dedicated displaced vertex analyses of these experiments provide bounds directly on the product $\sigma(A\to BX)\times\Gamma(X\to C)$ as a function of both $m_X$ and $c\tau_X$.
    \item \textbf{Invisible decays:} The new particle is produced in $A\to BX$ processes inside the detector, but it is very long-lived and decays outside the detector. The probability for a decay outside the detector is
    \begin{equation}
        P_\mathrm{out} = \exp\left(\frac{-L_\mathrm{max}}{c\tau_X \beta_X \gamma_X}\right)\,,
    \end{equation}
    where $L_\mathrm{max}$ is the maximum distance that the experiment is able to resolve. The experimental signature is a process with missing energy and momentum, which usually is interpreted as an excess in a process with neutrinos in the final state,
    \begin{equation}
        \sigma(A\to B +\mathrm{inv})_X = \sigma(A \to B X) \times P_\mathrm{out}.
    \end{equation}
    Another possible way to obtain an invisible decay, that we will not explore in this work, happens if the new particle acts as a portal to an invisible sector and decays predominantly to non-interacting dark matter particles.
\end{itemize}

From a purely computational point of view, the first difficulty that we encounter is the dependence of the predictions on $c\tau_X$, that forces us to calculate every decay channel of $X$ even for channels that are not directly observed. The exponential functions also induce non-Gaussianities in the likelihood function. Moreover, since some observables are only defined for certain values of the mass and/or the proper decay length of the particle, the likelihood function will not even be continuous at some points of the parameter space. All of this, together with the large number of model parameters and experimental observations at our disposal, make ML-based exploration particularly compelling.

\subsubsection{Axion-like particles}

One of the most popular and theoretically motivated examples of SM extensions featuring new light, feebly interacting particles are axions and axion-like particles (ALPs). Originally, axions were proposed as an elegant solution to the strong CP problem~\cite{Peccei:1977hh,Weinberg:1977ma,Wilczek:1977pj}, where a pseudo-scalar particle $a$ of mass $m_a$ and decay constant $f_a$, acting as the pseudo-Nambu Goldstone boson (pNGB) of the Peccei-Quinn (PQ) $U(1)_\mathrm{PQ}$ symmetry, dynamically drives the value of the electric dipole moment of the neutron to zero. However, the axion solution requires a correlation between its mass and decay constant, namely~\cite{Weinberg:1977ma,GrillidiCortona:2015jxo}
\begin{equation}
    m_a^2 f_a^2 = \frac{m_u m_d}{(m_u+m_d)^2} m_\pi^2 f_\pi^2\,,
\end{equation}
which severely constrains the available parameter space. Alternatively, if one chooses to consider $m_a$ and $f_a$ as independent parameters, we are instead in the ALP paradigm. While an ALP might not necessarily solve the strong CP problem, it has other interesting applications~\cite{Albertus:2026fbe}, like serving as a dark matter candidate~\cite{Abbott:1982af,Dine:1982ah,Preskill:1982cy,Co:2019jts,Chang:2019tvx,Co:2020dya}, explaining the flavour puzzle (axiflavon/flaxion~\cite{Ema:2016ops,Calibbi:2016hwq}), neutrino masses (majoron~\cite{Chikashige:1980qk,Chikashige:1980ui,Gelmini:1980re,Rothstein:1992rh}), or being generic predictions of string theory (axiverse~\cite{Arvanitaki:2009fg,Gendler:2023kjt}).

From a purely bottom-up perspective, we can study ALPs in the framework of EFTs~\cite{Georgi:1986df}, as an expansion in powers of $1/f_a$. The first terms of this expansion are the renormalisable operators for the kinetic and mass terms, and dimension-5 operators describing the lowest-order interactions with SM particles: derivative interactions with fermions that respect the shift-symmetry $a\to a + \mathrm{constant}$, and anomalous interactions with gauge bosons:
\begin{equation}
\begin{split}
    \mathcal{L}_\mathrm{ALP-EFT}^{d\leq 5} =& \mathcal{L}_\mathrm{SM} + \frac{1}{2}\partial_\mu a\,\partial^\mu a -\frac{1}{2}m_a^2 a^2 + \sum_{V=B, W, G} \frac{\alpha_V}{4\pi} \frac{c_V a}{f_a} V_{\mu\nu} \widetilde{V}^{\mu\nu}\\ &+\frac{\partial_\mu a}{f_a}\left[c_{q_L}^{ij} \bar q_i \gamma^\mu q_j + c_{u_R}^{ij}\bar u_i \gamma^\mu u_j + c_{d_R}^{ij}\bar d_i \gamma^\mu d_j + c_{\ell_R}^{ij}\bar \ell_i \gamma^\mu \ell_j + c_{e_R}^{ij}\bar e_i \gamma^\mu e_j \right]\,,
\end{split}
\end{equation}
where $\widetilde{V}^{\mu\nu} = \frac{1}{2}\epsilon^{\mu\nu\rho\sigma}V_{\rho\sigma}$ is the dual tensor in the convention $\epsilon^{0123}=+1$, $q$ and $\ell$ are the $SU(2)_L$ doublets of left-handed quarks and leptons, and $u$, $d$ and $e$ are the $SU(2)_L$ singlets of right-handed quarks and charged leptons. In general, the couplings to fermions $c_f^{ij}$ are hermitian $3\times3$ matrices in flavour space.

From the model building perspective, we can distinguish two very broad categories of UV-complete models, depending on which fermions are charged under the PQ symmetry. If only new heavy fermions are charged under PQ, as is the case in the KSVZ model~\cite{Kim:1979if,Shifman:1979if} and its generalizations~\cite{DiLuzio:2020wdo,Plakkot:2021xyx,DiLuzio:2024xnt}, the ALP at the matching scale $\Lambda$ will only interact with the SM gauge bosons as dictated by the gauge anomalies, unless the heavy and SM fermions mix. On the other hand, if the SM fermions $f_i$ have PQ charges $\mathcal{X}_f^i$, then the ALP at the matching scale will feature diagonal (but not necessarily universal) couplings to fermions $c_f^{ii}(\Lambda) = -\mathcal{X}_f^i$, and couplings to gauge bosons fixed by the anomalies~\cite{DiLuzio:2020wdo},
\begin{equation}\begin{split}
    c_G(\Lambda) &= -\frac{1}{2}\sum_i (2\mathcal{X}_{q_L}^i - \mathcal{X}_{u_R}^i - \mathcal{X}_{d_R}^i)\,,\\
    c_W(\Lambda) &= -\frac{1}{2}\sum_i (3\mathcal{X}_{q_L}^i + \mathcal{X}_{\ell_L}^i)\,,\\
    c_B(\Lambda) &= -\frac{1}{6}\sum_i (\mathcal{X}_{q_L}^i - 8\mathcal{X}_{u_R}^i-2\mathcal{X}_{d_R}^i + 3\mathcal{X}_{\ell_L}^i-6\mathcal{X}_{e_R}^i)\,.
\end{split}\end{equation}

Note that in specific realizations, like DFSZ model~\cite{Zhitnitsky:1980tq,Dine:1981rt} and its generalizations~\cite{DiLuzio:2020wdo,Diehl:2023uui,Cox:2023squ}, or flaxion models, the scalar sector will impose constraints to combinations of the PQ charges.

Once the ALP couplings have been determined at the matching scale $\Lambda \sim 4\pi f_a$, we must solve the Renormalization Group Equations (RGEs) and integrate out the heavy SM particles ($t$, $h$, $Z$, $W^\pm$) in order to obtain the values of the coefficients at the physical scale of each observable~\cite{Choi:2017gpf,MartinCamalich:2020dfe,Chala:2020wvs,Bauer:2020jbp,Bonilla:2021ufe,DasBakshi:2023lca,Bresciani:2024shu}. One consequence of the running and matching is that in models with UV couplings to top quarks and/or $W^\pm$, sizable off-diagonal couplings in the down sector will be generated.

\subsubsection{ALPs as a solution to the Belle II anomaly}

The Belle II experiment has found an excess in the branching ratio of $B^+\to K^+ \nu \bar\nu$ of $2.8\,\sigma$ with respect to the SM prediction~\cite{Belle-II:2023esi}. Several analyses of the $q^2$ distribution~\cite{Altmannshofer:2023hkn,Fridell:2023ssf,Bolton:2025fsq,Abumusabh:2025zsr,Gartner:2026clx} have shown that the excess is compatible with the production of one single particle of mass around $2\,\mathrm{GeV}$ that does not decay inside the detector. Furthermore, by combining also the upper limit for the process $B\to K^*\nu \bar \nu$ at BaBar~\cite{BaBar:2013npw}, with different detector size and Lorentz boost, Ref.~\cite{Alda:2025uwo} found a lower bound for the proper decay length of the particle, $c\tau \geq 80\,\mathrm{cm}\ (95\%\ \mathrm{C.L.})$.

The identification of this presumptive $2\,\mathrm{GeV}$ particle with an ALP proves to be challenging when we consider scenarios motivated by UV-complete models. On the one hand, the flavour-violating coupling $|c_{sb}^V|/f_a$ that enters in the ALP production $B^+ \to K^+ a$ needs to be large enough to explain the excess, but on the other hand, ALP decay channels such as $a \to \mu^+ \mu^-$ and $a \to \eta \pi \pi$ need to be suppressed in order to respect the constraint on the ALP lifetime.

Ref.~\cite{Alda:2025uwo} identified a model that fulfills all the requirements, based on the astrophobic axions of Refs.~\cite{DiLuzio:2017ogq,Bjorkeroth:2019jtx,DiLuzio:2022tyc}. The PQ charges of the SM fermions are
\begin{equation}
    \mathcal{X}_{q_L} = \mathcal{X}_{\ell_L} = (0,0,-1),\qquad \mathcal{X}_{u_R} = -\sin^2\beta\,,\qquad \mathcal{X}_{d_R} = \mathcal{X}_{e_R}=-\cos^2\beta\,.
\end{equation}

In this construction, the coupling $c_{sb}^V$ is generated by top loops, and it is largely independent of $\beta$. Meanwhile, the coupling to muons receives a tree level contribution $c_{\mu\mu}^{A(0)} = \cos^2\beta$ which is partially canceled by the running of the coupling to $\tau$ leptons $c_{\tau\tau}^A = \cos^2\beta-1 = -\sin^2 \beta$, and consequently it is minimized for intermediate $\beta$ angles.

\subsubsection{Fit parameters and results}

The model in~\cite{Alda:2025uwo} is rather predictive, as it only depends on two free parameters, $\beta$ and $f_a$. Here we aim to extend it in a more descriptive direction, exploring a larger region of the parameter space by lifting the dependence relations between the PQ charges. The free parameters of our new model will be the PQ charges of the third-generation left-handed fermions $\mathcal{X}_{q_L}^3$ and $\mathcal{X}_{\ell_L}^3$ and the universal PQ charges of the right-handed fermions $\mathcal{X}_{u_R}$, $\mathcal{X}_{d_R}$ and $\mathcal{X}_{e_R}$, all of them allowed in the range $\mathcal{X}_f \in [-1, 1]$, as well as the energy scale $f_a$ in the range $f_a \in [10^5, 10^8]\,\mathrm{GeV}$ (while fixing $\Lambda = 4\pi f_a$) and the ALP mass in the range $m_a \in [1.7,2.2]\,\mathrm{GeV}$.

We perform the fit of this model using the likelihood function implemented by the \texttt{ALP-aca} library~\cite{Alda:2025nsz}. This likelihood function includes more than 100 measurements, but not all of them are operative depending on the ALP mass and proper lifetime. In our setup, in addition to $B \to K^{(*)}a(\to  \mathrm{inv})$ processes, we have to consider off-shell ALP processes (leptonic and radiative meson decays, neutral meson mixing), and in the limit of short lifetime, prompt/displaced vertex processes like $B\to K^{(*)} a(\to  \mu^+ \mu^-)$ decays.

\begin{figure}
    \centering
    \includegraphics[width=0.5\linewidth]{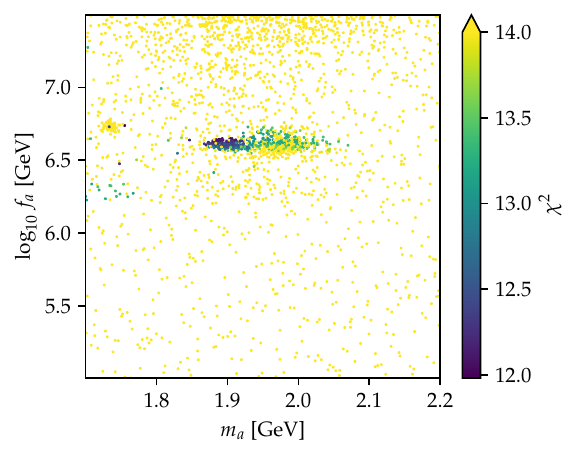}
    \caption{Points of the training dataset in the plane $m_a$-$\log_{10}f_a$. The active learning algorithm ensures that the density of points is larger in the areas of small $\chi^2$ (darker colours), while still covering the whole parameter space.}
    \label{fig:activelearning}
\end{figure}

The training dataset is generated following an active learning programme in order to improve the coverage of the region of large likelihood. A small sample (800 points) generated by LHS is used as a starting seed. Then, at each iteration of the active learning algorithm, a Gaussian Process (GP), the infinite-dimensional limit of a multivariate Gaussian, is trained on the dataset up to that point. A set of candidate points are proposed, for which the GP predicts its $\chi^2$, and more importantly, the uncertainty of said prediction. The candidate points are scored according to their modified Expected Improvement
\begin{equation}
    \mathrm{EI} = (\chi^2_* - \mu-\xi)\ \Phi\!\left(\frac{\chi^2_*-\mu-\xi}{\sigma}\right) + \sigma\ \phi\!\left(\frac{\chi^2_*-\mu-\xi}{\sigma}\right)\,,
\end{equation}
where $\mu$ and $\sigma$ are the predicted central value and uncertainty of the $\chi^2$ for the point, $\chi^2_*$ is the minimum $\chi^2$ observed so far, and $\Phi$ and $\phi$ are the cumulative distribution function and probability distribution function of the Gaussian distribution, respectively. The parameter $\xi$ regulates the trade-off between exploration and exploitation: if $\xi$ is large, the best score will be achieved by points where the GP is very uncertain, while for small $\xi$, points with small predicted $\chi^2$ are selected instead. At each iteration of the algorithm, the point with the best score is selected, its true $\chi^2$ is computed and added to the training dataset, and $\xi$ is slightly decreased. The total size of the dataset, including both the initial seeds and the points selected by the GP, is 2500 points. The training dataset is depicted in Figure~\ref{fig:activelearning}, showing that the highest density is achieved in the region of low $\chi^2$.

In order to fully capture the details of the $\chi^2$ function in the region around its minimum, we propose a two-step strategy: we first train a \texttt{XGBoost} instance on the sigmoid of the displaced $\chi^2$ function,
\begin{equation}
    \sigma(\chi^2) = \frac{1}{1+\exp\left(\frac{\chi^2 - \chi^2_c}{0.4}\right)}\,,
\end{equation}
where $\chi^2_c = \chi^2_\mathrm{b.f.} +10$ is the cutoff: this first stage acts as a classifier, retaining only the points below the cutoff. The second stage is a \texttt{XGBoost} instance trained only on those points below the cutoff, thus focusing on the relevant region of the parameter space.

\begin{figure}[H]
    
    \includegraphics[width=0.50\textwidth]{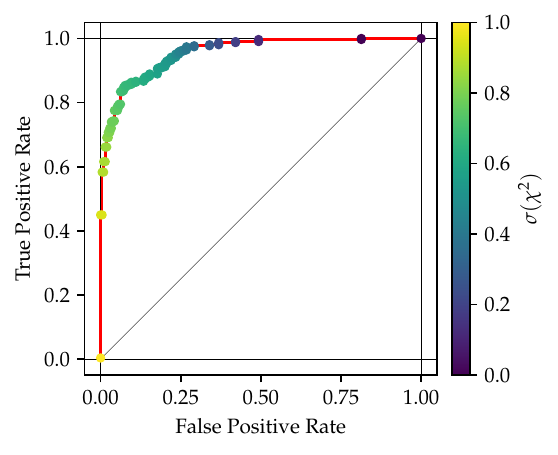} \includegraphics[width=0.42\textwidth]{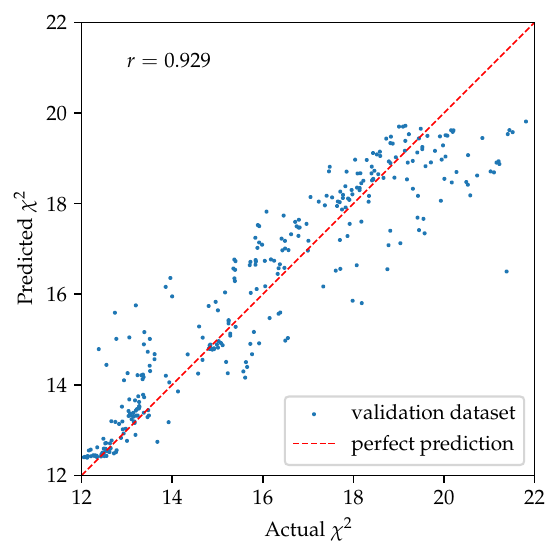}\\
    \makebox[0.50\textwidth]{(a)} \makebox[0.42\textwidth]{(b)}
    \caption{{(\textbf{a})} Receiver Operating Characteristic (ROC) curve of the first \texttt{XGBoost} trained on the ALP dataset. {(\textbf{b})} Predicted versus exact $\chi^2$ on the held-out validation dataset.}
    \label{fig:training_alps}
\end{figure}

The performance of the classifier stage is characterised by the receiver operating characteristic (ROC) curve in Figure~\ref{fig:training_alps}. It presents the true-positive rate (the proportion of points below the cutoff that the classifier correctly identifies) and the false-positive rate (the proportion of points above the cutoff that it misidentifies) for various thresholds of $\sigma(\chi^2)$. A random classifier would lie on the diagonal of the ROC curve, while a perfect classifier would be situated in the upper-left corner. Setting the classifier threshold to $\sigma(\chi^2)=0.5$ gives $\mathrm{TPR}=0.96$ and $\mathrm{FPR}=0.16$ on the validation dataset. The area under the ROC curve is $\mathrm{AUC} = 0.974$, indicating excellent classification performance. In the regressor stage, the surrogate achieves a Pearson correlation coefficient of $r=0.929$, as shown in Figure~\ref{fig:training_alps}b.

%\vspace{-9pt}

As in the $B$-physics case, the decisive advantage of the surrogate is its evaluation speed, quantified in Table~\ref{tab:alp_timing}.  On a single CPU core, one evaluation of the exact \texttt{ALP-aca} likelihood over the seven-dimensional parameter space takes $\sim\!1.9$\,s (it must solve the RGEs, perform the matching and recompute more than a hundred observables), whereas the trained XGBoost ensemble predicts the same point in $\sim\!3.5\,\upmu$s, a speed-up of about $5.3\times10^{5}$. The gain is most tangible in the construction of the training set: generating the $2.5\times10^{3}$ exact points costs about $1.3$ core-hours of \texttt{ALP-aca} computation, which the emulator re-evaluates in $\sim\!0.009$\,s. As in the flavour benchmark, all times are measured on a single core, so the comparison is reproducible and does not depend on the number of cores or on the specific machine.

\begin{table}[H]

\caption{Evaluation cost of the XGBoost surrogate against the exact \texttt{ALP-aca} likelihood, on a single CPU core.  As in the $B$-physics benchmark, per-evaluation times use $10^{4}$ surrogate predictions (batch) and a serial run of distinct exact evaluations; the second row is the cost of generating the training set.}
\label{tab:alp_timing}

\setlength{\tabcolsep}{5.3mm}
\resizebox{\textwidth}{!}{\begin{tabular}{lccc}
\toprule
 & \textbf{XGBoost Emulator} & \textbf{Exact \texttt{ALP-aca}} & \textbf{Speed-Up} \\
\midrule
Per evaluation  & $3.5\,\upmu$s & $1.85$\,s     & $5.3\times10^{5}$ \\
Training set ($2.5\times10^{3}$ points) & $0.009$\,s   & $1.3$\,core-h & $5.3\times10^{5}$ \\
\bottomrule
\end{tabular}}

\end{table}

\begin{figure}[H]
    
    \includegraphics[width=0.46\linewidth]{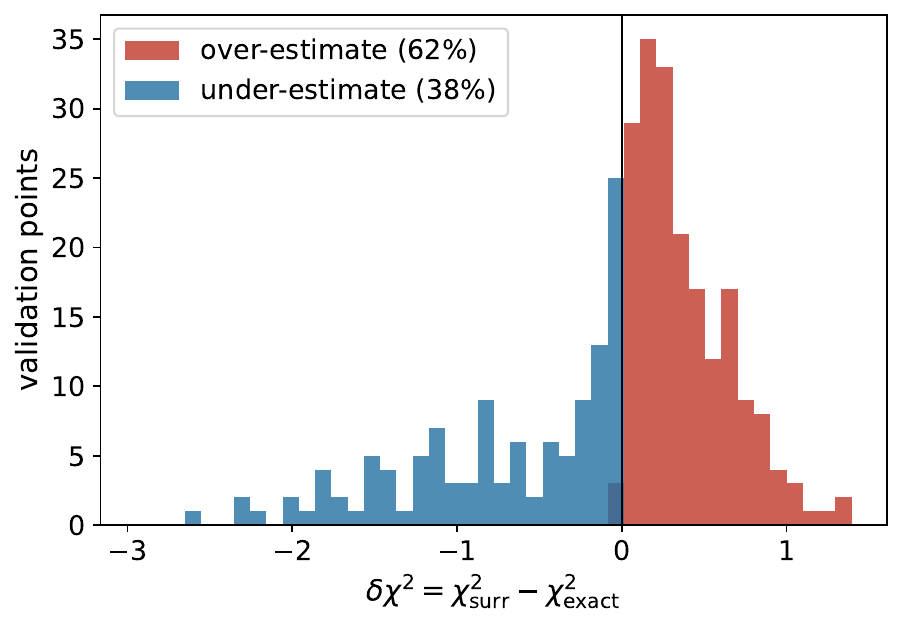}\hfill
    \includegraphics[width=0.46\linewidth]{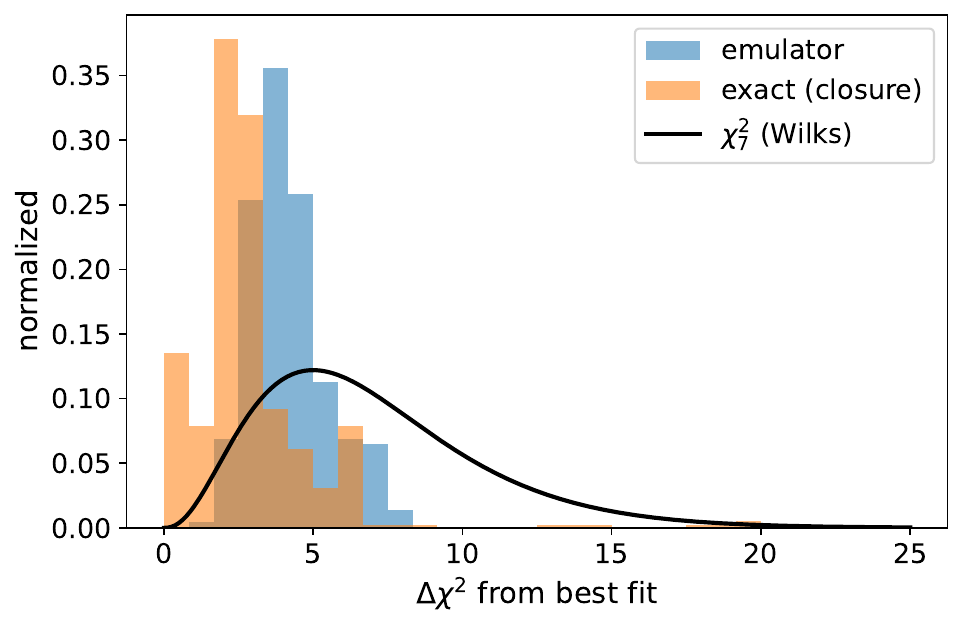}\\
    \makebox[0.46\linewidth]{(a)}\hfill\makebox[0.46\linewidth]{(b)}
    \caption{Validation of the ALP emulator.
    {(\textbf{a})} Signed residuals $\delta\chi^2=\chi^2_{\rm surr}-\chi^2_{\rm exact}$ on the held-out validation set, split into over- (red) and under-estimates (blue).
    {(\textbf{b})} Posterior-level closure test: distribution of $\Delta\chi^2=\chi^2-\chi^2_{\rm min}$ over a sample drawn from the emulated posterior, scored with the emulator (blue) and by re-evaluating the exact \texttt{ALP-aca} $\chi^2$ at the same points (orange), together with the $\chi^2$ distribution for seven degrees of freedom (Gaussian/Wilks limit, black).}
    \label{fig:alp_perf}
\end{figure}

Analogously to the flavour case, we validate the surrogate specifically in the low-$\Delta\chi^2$ region that sets the confidence contours, and at the level of the posterior it actually samples.  On a held-out validation split the signed residuals $\delta\chi^2=\chi^2_{\rm surr}-\chi^2_{\rm exact}$ are tightly concentrated around zero (Figure~\ref{fig:alp_perf}a) with a negligible mean bias ($-0.05$ in $\chi^2$ units): $62\%$ of the points are over-estimated by a small amount (mean $+0.40$) and $38\%$ under-estimated by a slightly larger amount (mean $-0.78$), so the two tails almost cancel and the regressor shows no systematic tendency to over- or under-cover.  Its mean absolute error, already small over the whole allowed region ($\mathrm{MAE}=0.54$), decreases further in the band that defines the contours, to $0.43$ within $2\,\sigma$ and $0.29$ within $1\,\sigma$ (Table~\ref{tab:alp_validation}).

\vspace{-6pt}
\begin{table}[H]
 
 \caption{Emulator error (in $\chi^2$ units) on the held-out ALP validation set, globally and restricted to the low-$\Delta\chi^2$ region that sets the confidence contours.}
\label{tab:alp_validation}

\setlength{\tabcolsep}{9.3mm}
\resizebox{\textwidth}{!}{\begin{tabular}{lccc}
\toprule
\textbf{Region (Exact \boldmath$\Delta\chi^2$ from Best Fit)} & \boldmath$N$ & \textbf{MAE} & \textbf{RMSE} \\
\midrule
all validation points   & 319 & 0.54 & 0.81 \\
within $2\,\sigma$ ($\Delta\chi^2\le6.18$)   & 240 & 0.43 & 0.60 \\
within $1\,\sigma$ ($\Delta\chi^2\le2.30$)   & 101 & 0.29 & 0.40 \\
\bottomrule
\end{tabular}}

\end{table}

We then perform a posterior-level closure test (Figure~\ref{fig:alp_perf}b).  Because the ALP posterior is a seven-dimensional distribution rather than a set of two-dimensional maps, and the exact likelihood cannot be tabulated on a dense grid, this closure test plays the same role as the direct comparison of exact and emulated likelihood maps (Figure~\ref{fig:bphys_maps}) in the flavour case: we draw a representative sample from the emulated posterior (Section~\ref{sec:ml}) and re-evaluate the exact \texttt{ALP-aca} $\chi^2$ at those same points.  Two features specific to this light-NP likelihood emerge.  First, the exact $\chi^2$ involves numerical integration using a Monte Carlo integrator (\texttt{vegas}) and therefore contains intrinsic statistical noise. As estimated by re-evaluating a subset with independent random seeds, this noise floor is small (RMS $\approx\!0.05$ in $\chi^2$ units for the numerically stable points), and the surrogate, trained on many such noisy labels, effectively averages out the noise and approximates the underlying noise-free $\chi^2$.

Second, and unlike the flavour case, a sizeable fraction ($\sim\!40\%$) of the points drawn from the emulated posterior return a large exact $\chi^2$. These large values are physical rather than numerical: they are orders of magnitude larger than the \texttt{vegas} noise floor quoted above, and their number is unchanged when the number of integrand evaluations is increased eightfold. They correspond to physically disfavoured points, at which the ALP would be visible and therefore excluded by the relevant experiments, so that the exact likelihood rises very steeply and assigns them a genuinely large $\chi^2$.  

For the points in the allowed region, the two $\Delta\chi^2$ distributions are visibly displaced relative to each other (Figure~\ref{fig:alp_perf}b): the emulator histogram lies at somewhat larger $\Delta\chi^2$ than the exact one, overestimating the exact $\chi^2$ by a mean offset of $+1.8$ over the posterior, with a root-mean-square difference of $\approx\!2.0$ in $\chi^2$ units within $2\,\sigma$.  This gap between the two~histograms is nearly constant across the region, changing by only $\approx\!0.2$ between the $1\,\sigma$ and $2\,\sigma$ bands. When referenced to each distribution's own minimum, it therefore cancels in $\Delta\chi^2=\chi^2-\chi^2_{\rm min}$, shifting the confidence contours by a comparably small amount in the conservative (over-covering) direction.

Beyond this mutual offset, both distributions lie at larger $\Delta\chi^2$ than the $\chi^2_7$ (Wilks) reference. This common displacement, shared by the emulator and the exact likelihood, is a genuine imprint of the non-Gaussian, partly discontinuous ALP likelihood and must be distinguished from the emulator bias, which is the relative shift between the two histograms.  The closure is therefore looser than in the flavour benchmark, as expected for a seven-dimensional, partly discontinuous likelihood emulated from a training set that is an order of magnitude smaller, but it confirms that, over the region occupied by the posterior, the surrogate reproduces the exact likelihood with sufficient accuracy not to affect the confidence regions.  As in the flavour application, the surrogate thus operates here not merely as an accelerator but as a statistically faithful replacement for the exact likelihood in the low-$\Delta\chi^2$ region that sets the ALP limits.

\begin{figure}[H]
     
   \hspace{-6pt} \includegraphics[width=0.48\linewidth]{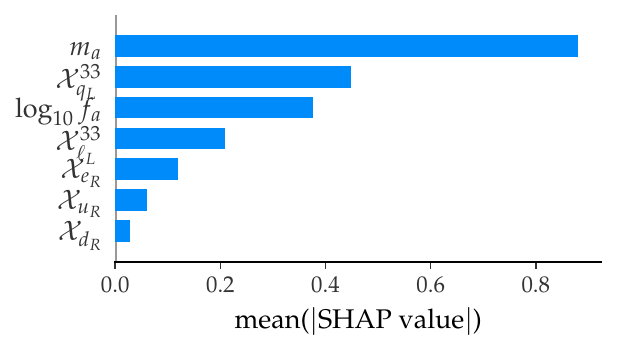} \includegraphics[width=0.40\linewidth]{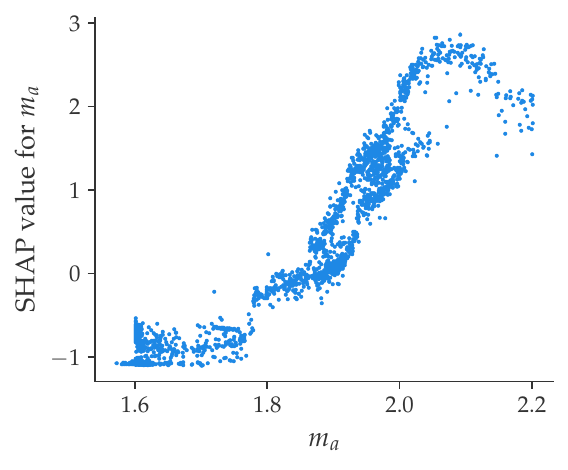}\\
    \hspace{-6pt} \makebox[0.48\linewidth]{(a)} \makebox[0.40\linewidth]{(b)}
    \caption{{(\textbf{a})} Global SHAP feature importance (mean absolute SHAP value, i.e.,\ each parameter's average contribution to the emulated $\chi^2$) for the parameters of the ALP fit. {(\textbf{b})} Point-wise SHAP values for $m_a$.}
    \label{fig:shapsummary_alps}
\end{figure}

The SHAP values help us again to interpret the results of the fit. In Figure~\ref{fig:shapsummary_alps} (left), we have displayed the mean of absolute values of SHAPs in the dataset, which corresponds to the expected importance of each parameter. The most important parameter is $m_a$, since it determines if a solution to the Belle II anomaly is even kinematically allowed in the first place. After that, the two most important parameters are $\mathcal{X}_{q_L}^{3}$ and $f_a$, which control the top loops and the overall impact of ALP effects. The impact of the PQ charges of right-handed fermions is more limited. In the case of $m_a$, we can see in Figure~\ref{fig:shapsummary_alps} (right) that larger ALP masses, which are less compatible with the $q^2$ distribution of the recast to two-body decay kinematics, result in larger SHAP values, that is, larger contribution to the $\chi^2$ function. 

\begin{figure}[H]
    
 \includegraphics[width=0.72\linewidth]{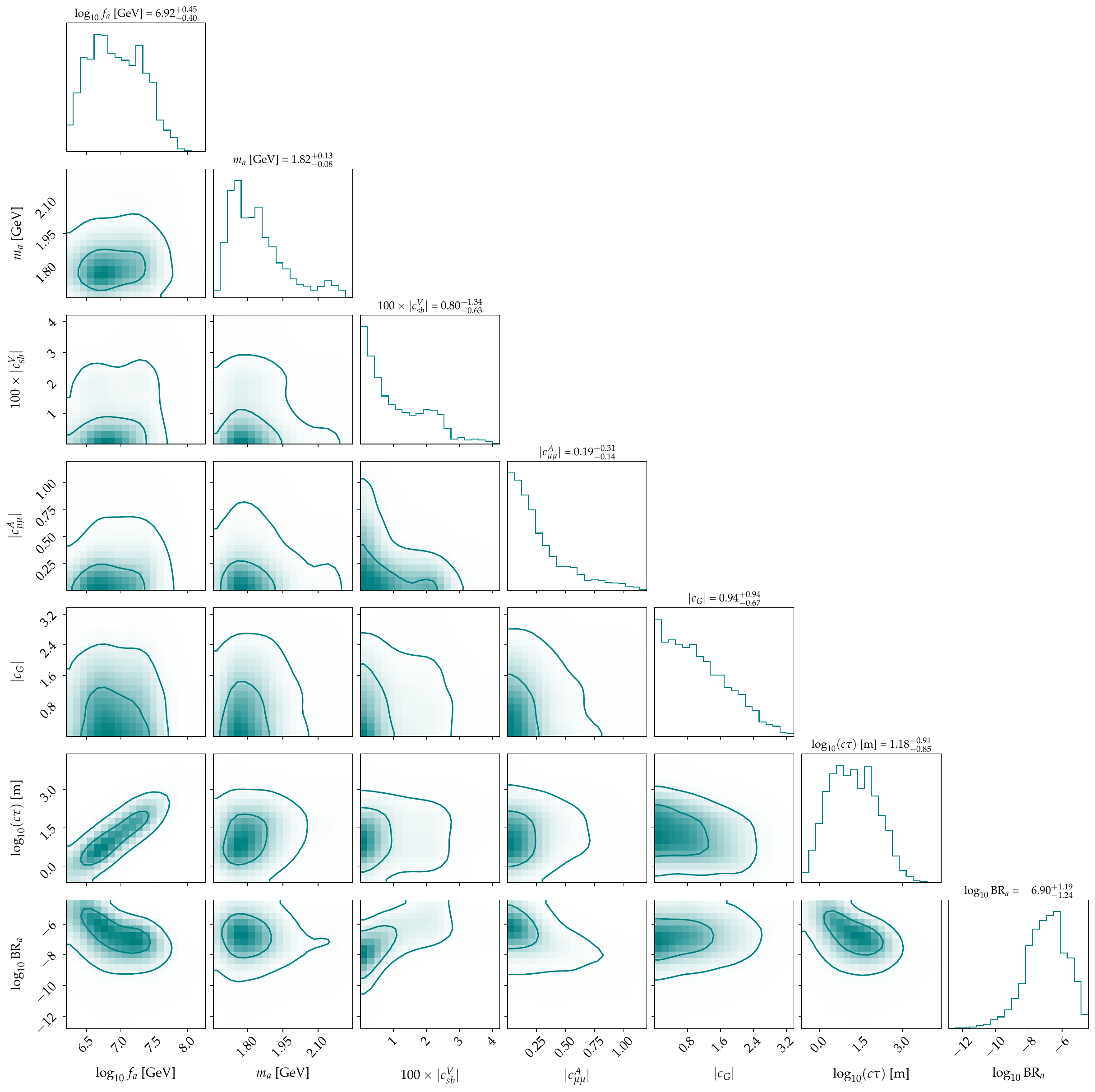}
 \centering
    \caption{Posterior distribution of ALP parameters and derived quantities. The contours correspond to the $1\,\sigma$ and $2\,\sigma$ regions, while the intensity of the shaded area to the density of posterior probability.}
    \label{fig:corner_alp}
\end{figure}

Finally, using the surrogate of the likelihood function and a Markov chain Monte Carlo (MCMC) implemented by the \texttt{emcee} library, we can efficiently sample the posterior distributions of both the parameters of the model and of derived quantities. In Figure~\ref{fig:corner_alp}, we have obtained the posterior distributions for the input parameters $m_a$ and $f_a$, for the low-energy couplings $|c_{sb}^V|$, $|c_{\mu\mu}^A|$ and $|c_G|$ (all of them at the physical scale $\mu\sim m_a$) and the relevant parameters for the Belle II anomaly: $c\tau_a$ and $\mathrm{BR}(B\to K a)$. The posterior distribution correctly reflects functional dependencies, such as $c\tau_a \propto f_a^2$ and $\mathrm{BR}(B^+\to K^+a) \propto |c_{sb}^V|^2$. Clearly, results favor an ALP mass around $1.8$ GeV, in agreement with previous analyses of the Belle II excess based on the kinematic reconstruction of the missing-energy signal. The preferred values of $f_a$ lie in the $10^{6}-10^{7}$ GeV range, balancing the need for sufficiently large flavour-changing interactions to explain the anomaly against the suppression required by constraints included in the fit. The posterior distributions of the effective couplings show that $|c_{sb}^V|^2$ must be large enough to generate an observable $\mathrm{BR}(B^+\to K^+a)$, while $|c_{\mu\mu}^A|$ remains comparatively small in order to suppress visible ALP decays and ensure compatibility with searches for displaced and prompt dimuon resonances.

\begin{figure}[H]
    \centering
    \includegraphics[width=0.65\linewidth]{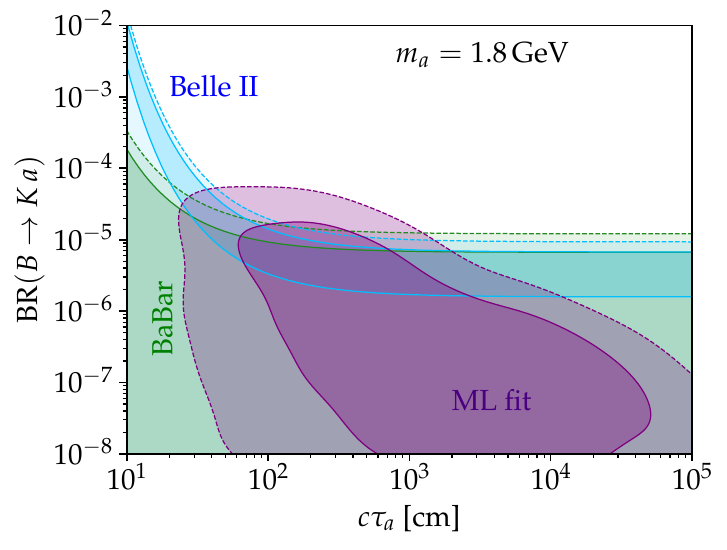}
    \caption{Allowed region in the $c\tau_a - \mathrm{BR}(B\to K a)$ plane determined by the surrogate-based fit including all observables (purple) compared to the allowed regions by $\mathrm{BR}(B\to K +\mathrm{inv})$ alone (BaBar in green, Belle II in blue). Lighter and darker shades correspond to $1\,\sigma$ and $2\,\sigma$ regions, respectively. The ALP mass is fixed to its best fit value.}
    \label{fig:fit_BKa}
\end{figure}

Figure~\ref{fig:fit_BKa} again shows the posterior obtained with the surrogate and MCMC, this time for $c\tau_a$ and $\mathrm{BR}(B^+\to K^+ a)$ (in purple) with $m_a=1.8\,\mathrm{GeV}$ fixed, compared with the fit using the $B\to K +\mathrm{inv}$ data alone. The BaBar and Belle~II data can set only a lower bound on the ALP decay length, but their combination with other observables in the fit provides complementary information that further constrains the proper decay length. As in the SMEFT analysis, this result underlines the importance of global explorations of parameter space, which benefit greatly from ML techniques such as those described here.

\subsection{Cosmology}\label{sec:cosmo}

In recent decades, cosmology has become a precision science, and its methods of parameter inference and data analysis have grown increasingly similar to those used in high-energy physics. Cosmological parameters can be inferred using several types of probes. The geometry of space-time can be determined from the distance--redshift relation using time-domain probes or by studying the large-scale structure of the Universe. These probes have enabled us to define a baseline cosmological model, flat $\Lambda$CDM, but tensions among observations may indicate the need for NP~\cite{CosmoVerseNetwork:2025alb}. In each case, the likelihood is again built from a $\chi^2$ that compares a handful of cosmological parameters with hundreds of measurements through theoretical predictions requiring numerical distance integrals. The same obstacles encountered in the flavour fits---expensive forward evaluations and curved parameter degeneracies---therefore reappear here, making the ML emulation strategy of Section~\ref{sec:ml} equally attractive. The main cosmological probes used in this work are the distance--redshift relation, measured using Type~Ia supernovae as standard candles and baryon acoustic oscillations as a standard ruler, complemented by the cosmic microwave background (CMB) as a prior on cosmological parameters.

\subsubsection{Cosmological probes and the likelihood}
\label{sec:cosmo_probes}

The baseline model for cosmological analysis is the flat $\Lambda$CDM model where the accelerated expansion of the Universe is explained by the cosmological $\Lambda$ with constant density and spatial flatness at each moment of cosmological time. The expansion history of the Universe is encoded in the dimensionless Friedmann equation:
$E(z)\equiv H(z)/H_0$,
\begin{equation}
    E(z) = \Big[\,\Omega_m (1+z)^3
        + (1-\Omega_m)\,f_{DE}(z)\,\Big]^{1/2},
    \label{eq:Ez}
\end{equation}
where spatial flatness fixes the dark-energy density parameter to $1-\Omega_m$ and radiation is negligible at the redshifts probed. The dynamics are driven by the parameters: $H_0$ is the Hubble constant, $\Omega_m=\Omega_b+\Omega_{cdm}$ is the density parameter of matter in the Universe, comprised by ordinary matter ($\Omega_b$) and cold dark matter ($\Omega_{cdm}$). The dynamical nature of dark energy is enclosed in $f_{DE}(z)$. For the $\Lambda$CDM model, as the energy density of the cosmological constant component does not evolve with time ($\rho_{\Lambda}\sim const.$), $f_{DE}=1$.

Over the past decade, various established and new cosmological probes have converged on $\Lambda$CDM as the most plausible cosmological model. More recently, however, supernova and standard-ruler measurements have indicated a possible dynamical-dark-energy component. The evidence for dynamical dark energy is assessed using the flat $w_0w_a$CDM model, in which the dark-energy equation of state is allowed to evolve with redshift according to the Chevallier--Polarski--Linder (CPL) parameterisation~\cite{Chevallier:2000qy,Linder:2002et},
\begin{equation}
    w(z) = w_0 + w_a\,\frac{z}{1+z}\,,
    \label{eq:cpl}
\end{equation}
which reduces to a cosmological constant ($\Lambda$CDM) for $w_0=-1$, $w_a=0$. In this particular case, the energy density of dynamical dark energy component evolves according to:
\begin{equation}
     f_{DE}(z) = (1+z)^{3(1+w_0+w_a)} e^{-3 w_a z/(1+z)}.
    \label{eq:fEz}
\end{equation}

All geometric observables follow from common definitions. The transverse distance between two objects at a given redshift is the comoving (transverse) distance, $D_M(z)$, while the radial geometric distance relation with redshift is the Hubble distance, $D_H(z)$, and finally we can capture the isotropic nature of a spherical ruler with the spherically averaged distance, $D_V(z)$,~\cite{SDSS:2005xqv}
\begin{equation}
    D_M(z) = \frac{c}{H_0}\int_0^{z}\frac{dz'}{E(z')}\,,\quad
    D_H(z) = \frac{c}{H(z)} = \frac{c}{H_0\,E(z)}\,,\quad
    D_V(z) = \big[\,z\,D_M^2(z)\,D_H(z)\,\big]^{1/3}\,,
    \label{eq:distances}
\end{equation}
so that the model depends on the four parameters $\bm{\theta}=(\Omega_m, H_0, w_0, w_a)$ (reduced to $(\Omega_m, H_0)$ in the $\Lambda$CDM baseline below).  The three probes enter the global likelihood as follows:
\begin{itemize}
\item \textbf{Type~Ia Supernovae} are standardisable candles: after light-curve correction their peak brightness is fixed, so the observed apparent magnitude $m_B$ measures the luminosity distance through the distance modulus
\begin{eqnarray}
    \mu(z) &\equiv& m_B - M = 5\log_{10}\!\big[d_L(z)/10\,\mathrm{pc}\big]
          = 5\log_{10}\!\big(d_L(z)/\mathrm{Mpc}\big) + 25\,,\nonumber\\
    d_L(z) &=& (1+z)\,D_M(z)\,,
    \label{eq:mu}
\end{eqnarray}
with $M$ the absolute magnitude.  Because $M$ is fully degenerate with $H_0$, it is analytically marginalised, so the supernovae constrain the shape of the distance--redshift relation rather than its absolute scale.  The $\chi^2$ uses the full statistical$+$systematic covariance matrix $\mathcal{C}$, $\chi^2_{\rm SN}=\Delta^{\mathsf T}\mathcal{C}^{-1}\Delta$ with $\Delta_i=\mu_i^{\rm obs}-\mu(z_i)$.  We use the Pantheon$+$\,/\,SH0ES compilation~\cite{Scolnic:2021amr,Brout:2022vxf} (1657 supernovae, with the Cepheid calibrators excluded so that $M$ is marginalised) and, as an independent sample, the Dark Energy Survey five-year SN\,Ia data set (DES-SN5YR), in both its 2024 release (1829 supernovae)~\cite{DES:2024tys} and its 2025 update~\cite{DES:2025sn5yr} (1820 supernovae); for DES the magnitude offset is marginalised analytically, including the corresponding normalisation term.

\item \textbf{Baryon Acoustic Oscillations} imprint a fixed comoving scale, the sound horizon at the baryon drag epoch $r_d$, that acts as a standard ruler in the galaxy distribution. The sound horizon is given by: $r_d=\int_{z_d}^{\infty}{\frac{c_s(z)}{H(z)}dz}$, where $z_d$ is the redshift at decoupling and the speed of sound prior to recombination, $c_s(z)$, is set by the baryon, $\rho_b$ and radiation, $\rho_\gamma$, densities as $c_s(z)=c\left[ \sqrt{3\left(1+\frac{3\rho_b(z)}{4\rho_\gamma(z)}\right)} \right]^{-1}$. Surveys measure the ratios $D_M/r_d$, $D_H/r_d$ and the isotropic combination $D_V/r_d$ at several effective redshifts.  We use the 13 measurements of the DESI Data Release~2~\cite{DESI:2025zgx} with their Gaussian likelihood $\chi^2_{\rm BAO}=\Delta^{\mathsf T}\mathcal{C}^{-1}_{\rm BAO}\Delta$, fixing $r_d$ to its Planck fiducial value.

\item \textbf{Cosmic Microwave Background.}  Rather than recomputing the full CMB power spectra at every point, we summarise the Planck constraint as a Gaussian prior on the late-time parameters. The Planck PR4$+$lensing chain in the CPL model obtained in the re-analysis of the latest Planck data~\cite{Tristram:2023haj} (superseding the official PR3~\cite{Planck:2018vyg}) made by~\cite{2026PDU....5202283A} provides a covariance matrix for its full set of about $25$ parameters; of these we keep only the five that map onto our fit, the Hubble constant $H_0$, the two physical densities of baryons $\omega_b$ and cold dark matter $\omega_{\rm cdm}$ (with $\omega_x\equiv\Omega_x h^2$ and $h=H_0/100$), and the dark-energy parameters $w_0$ and $w_a$, discarding the remaining nuisance columns. Because our sampling variable is the total matter density rather than its two components, we combine them into $\omega_m=\omega_b+\omega_{\rm cdm}$ by linear error propagation, which collapses the $3\times3$ block of $(H_0,\omega_b,\omega_{\rm cdm})$ into a correlated $2\times2$ covariance for $(H_0,\omega_m)$; the dark-energy parameters are carried with their Planck variances. The result is a $4\times4$ covariance $\Sigma_{\rm CMB}$ for $(H_0,\omega_m,w_0,w_a)$ in which $H_0$ and $\omega_m$ stay correlated, while $w_0$ and $w_a$ enter as independent Gaussian constraints.  Centred on the Planck $\Lambda$CDM best fit ($w_0=-1$, $w_a=0$), it contributes
\begin{equation}
    \chi^2_{\rm CMB} = \Delta^{\mathsf T}\,\Sigma_{\rm CMB}^{-1}\,\Delta\,,\qquad
    \Delta = \big(H_0-H_0^\star,\ \omega_m-\omega_m^\star,\
             w_0-w_0^\star,\ w_a-w_a^\star\big)\,,
    \label{eq:chi2_cmb}
\end{equation}
where $\omega_m=\Omega_m (H_0/100)^2$ is evaluated from the sampled parameters and the starred quantities are the Planck central values.  Combining the two densities by propagating their covariance is the exact marginalisation of the joint Gaussian onto $(H_0,\omega_m)$, so for observables that depend only on the total matter density it is equivalent to carrying $\omega_b$ and $\omega_{\rm cdm}$ as explicit parameters; the correlated $(H_0,\omega_m)$ block is retained, while in this particular covariance $w_0$ and $w_a$ happen to be uncorrelated with the rest and enter as independent Gaussians of width $\sigma(w_0)=0.02$, $\sigma(w_a)=0.05$. The same prior is used in every CMB combination of Sections~\ref{sec:cosmo_w0wa} and~\ref{sec:cosmo_bench}.
\end{itemize}

Combined data sets are fit by summing the individual $\chi^2$ functions, $\chi^2_{\rm tot}(\bm{\theta})=\sum_k \chi^2_k(\bm{\theta})$, exactly as in the flavour global fit.

\subsubsection{Surrogate construction and the choice of training target}\label{sec:cosmo_ml}

For each data combination we train an XGBoost surrogate of $\chi^2_{\rm tot}(\bm{\theta})$ following the methodology of Section~\ref{sec:ml}.  The training design combines two-dimensional slices through the best-fit point, a uniform space-filling draw over the full prior box ($\Omega_m\!\in\![0.1,0.9]$, $H_0\!\in\![20,100]$, $w_0\!\in\![-3,0.2]$, $w_a\!\in\![-3,2]$), and a Gaussian cloud whose covariance is the Hessian of the fit, so that the narrow, anisotropic $w_0$--$w_a$ degeneracy is sampled densely where it matters most.  Each design contains $2.5$--$5\times10^{5}$ points, with the exact distance integrals of Eqs.~(\ref{eq:fEz})--(\ref{eq:distances}) evaluated in parallel across CPU cores.

A central, problem-specific ingredient is the choice of training target. The cosmological $\chi^2$ spans an enormous dynamic range, from $\chi^2_{\rm min}\!\sim\!10^{3}$ at the best fit to $\gtrsim\!10^{6}$ at the edges of the prior box, while the physically relevant structure, the $\Delta\chi^2\!\sim\!2$--$6$ band that defines the $1\,\sigma$ and $2\,\sigma$ confidence regions, lives in a tiny sliver near the minimum.  Crucially, the absolute scale of $\chi^2_{\rm min}$ differs by two orders of magnitude between the data combinations, from $\chi^2_{\rm min}\!\approx\!1.7\times10^{3}$ for the supernova$+$BAO fits (more than $1600$ data points) down to $\chi^2_{\rm min}\!\approx\!5.6$ for the BAO-only fit (13 points), and this is what makes the two naive targets inadequate in a data-dependent way:
\begin{itemize}
    \item Training directly on $\chi^2$ lets the loss be dominated by the high-$\chi^2$ tail: the surrogate fits the outer wall of the likelihood well but is blind to the $\mathcal{O}(\text{few})$ variations near the minimum. This happens for every combination, irrespective of $\chi^2_{\rm min}$, and makes the recovered contours unusable.
    \item Training on a plain logarithm, $\log_{10}\chi^2$, compresses the tail, but whether it works now depends on where the minimum falls on the logarithmic curve, which is precisely why it succeeds for some data sets and fails for others.  For the supernova$+$BAO and $+$CMB fits the minimum is large, so it sits on the locally flat part of the logarithm and the decisive $1\,\sigma$--$2\,\sigma$ band maps onto a negligible change of the target ($\log_{10}1706-\log_{10}1700\approx1.5\times10^{-3}$): the contours remain unresolved. For the BAO-only fit the minimum is small, so it sits on the steep part of the logarithm ($\log_{10}12-\log_{10}5.6\approx0.33$) and the plain transform does resolve the contours.  No single un-shifted logarithm can therefore serve all combinations at once.
\end{itemize}

Both problems are solved by a shifted-$\log_{10}$ transform, which is the
target adopted throughout this work:
\begin{equation}
    y(\bm{\theta}) = \log_{10}\!\big(\chi^2(\bm{\theta}) - \chi^2_{\rm min} + 1\big)\,,
    \qquad
    \chi^2(\bm{\theta}) = 10^{\,y(\bm{\theta})} - 1 + \chi^2_{\rm min}\,,
    \label{eq:log_transform}
\end{equation}
where $\chi^2_{\rm min}$ is the minimum over the training set.  The shift maps the best fit to $y=0$ regardless of the absolute value of $\chi^2_{\rm min}$, so that the minimum of every likelihood, whether $\chi^2_{\rm min}\!\approx\!5$ or $\approx\!1700$, is placed on the same steep, well-resolved part of the transform; this is what removes the data-dependence of the plain logarithm discussed above. The unit offset additionally guarantees a strictly positive argument ($\geq1$), removing the $\log(0)$ singularity that a naive shift $\chi^2-\chi^2_{\rm min}$ would introduce exactly at the densely sampled best-fit point.  The transform compresses the $10^{6}$ tail and stretches precisely the near-minimum band that encodes the confidence regions, amplifying it by about three orders of magnitude in target space. The surrogate is trained on $y$ and its predictions are mapped back to linear $\chi^2$ through the exact inverse in Eq.~(\ref{eq:log_transform}), so that downstream contour and SHAP computations operate on physical $\chi^2$ values.  With this target the emulators reach a coefficient of determination $R^2=0.997$--$0.9999$ on a held-out validation set across all data combinations.

Posterior distributions are then mapped by running the parallel Random-Walk Metropolis--Hastings sampler of Section~\ref{sec:ml} (1024 chains, with the proposal covariance set to the Hessian of the fit) directly on the emulated likelihood, and rendered as marginalised contours. Because each likelihood call is a single forward pass through the tree ensemble rather than a set of numerical distance integrals, the cost of the posterior exploration is reduced by more than an order of magnitude, as quantified in Section~\ref{sec:cosmo_bench}.

\subsubsection{Results: The \texorpdfstring{$\Lambda$CDM}{Lambda-CDM} Baseline and Dynamical Dark Energy}
\label{sec:cosmo_w0wa}

\begin{figure}[H]
    \centering
    \includegraphics[width=0.45\linewidth]{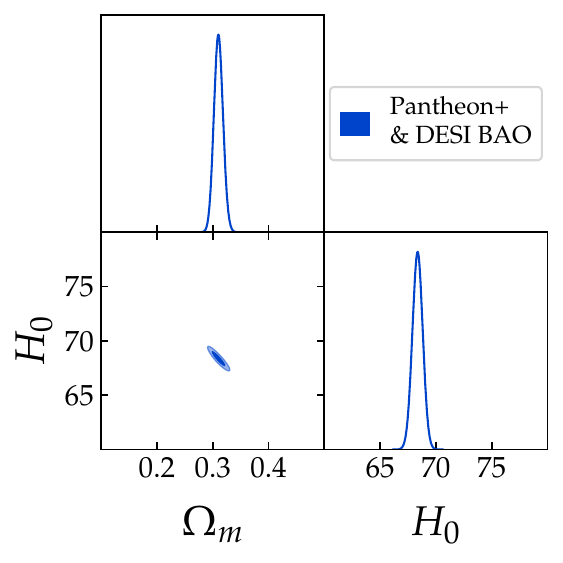}
    \caption{Marginalised $\Omega_m$--$H_0$ posterior of the $\Lambda$CDM baseline (Pantheon$+$\,$\&$\,DESI~BAO), obtained by sampling the XGBoost surrogate ($R^2=0.99988$) with the Random-Walk Metropolis--Hastings sampler. The inner and outer contours are the $1\,\sigma$ and $2\,\sigma$ credible regions.}
    \label{fig:cosmo_lcdm}
\end{figure}

As a simple, fully controlled example we first emulate the two-parameter
$\Lambda$CDM likelihood of Pantheon$+$ combined with DESI BAO, fitting only $(\Omega_m, H_0)$.  The surrogate reproduces the exact $\chi^2$ with $R^2=0.99988$, and the emulator MCMC recovers
\begin{equation}
    \Omega_m = 0.310 \pm 0.008\,,\qquad H_0 = 68.4 \pm 0.5~\mathrm{km\,s^{-1}Mpc^{-1}}\,,
\end{equation}
at $\chi^2_{\rm min}=1703.6$.  Figure~\ref{fig:cosmo_lcdm} shows the resulting $\Omega_m$--$H_0$ posterior: a single, closed, mildly anti-correlated ellipse with no secondary modes or open directions.  These values reproduce the standard concordance picture, $\Omega_m=0.310$ matching the Planck and DESI determinations to within $1\,\sigma$ and the supernova-only Pantheon$+$ value $\Omega_m\!\approx\!0.33$.  Because the supernova absolute magnitude is marginalised, the fit carries no SH0ES distance-ladder anchor, so $H_0$ is set by the BAO scale with the Planck sound horizon $r_d$; it therefore lands on the Planck-like value $H_0\!\approx\!68$ rather than the higher local-distance-ladder measurement, exactly as this dataset combination behaves in the DESI and Pantheon$+$ analyses. This baseline confirms that the pipeline reproduces the established result before turning to the more demanding, and more interesting, four-parameter dark-energy case.

\begin{figure}[H]
     
    \includegraphics[width=0.49\linewidth]{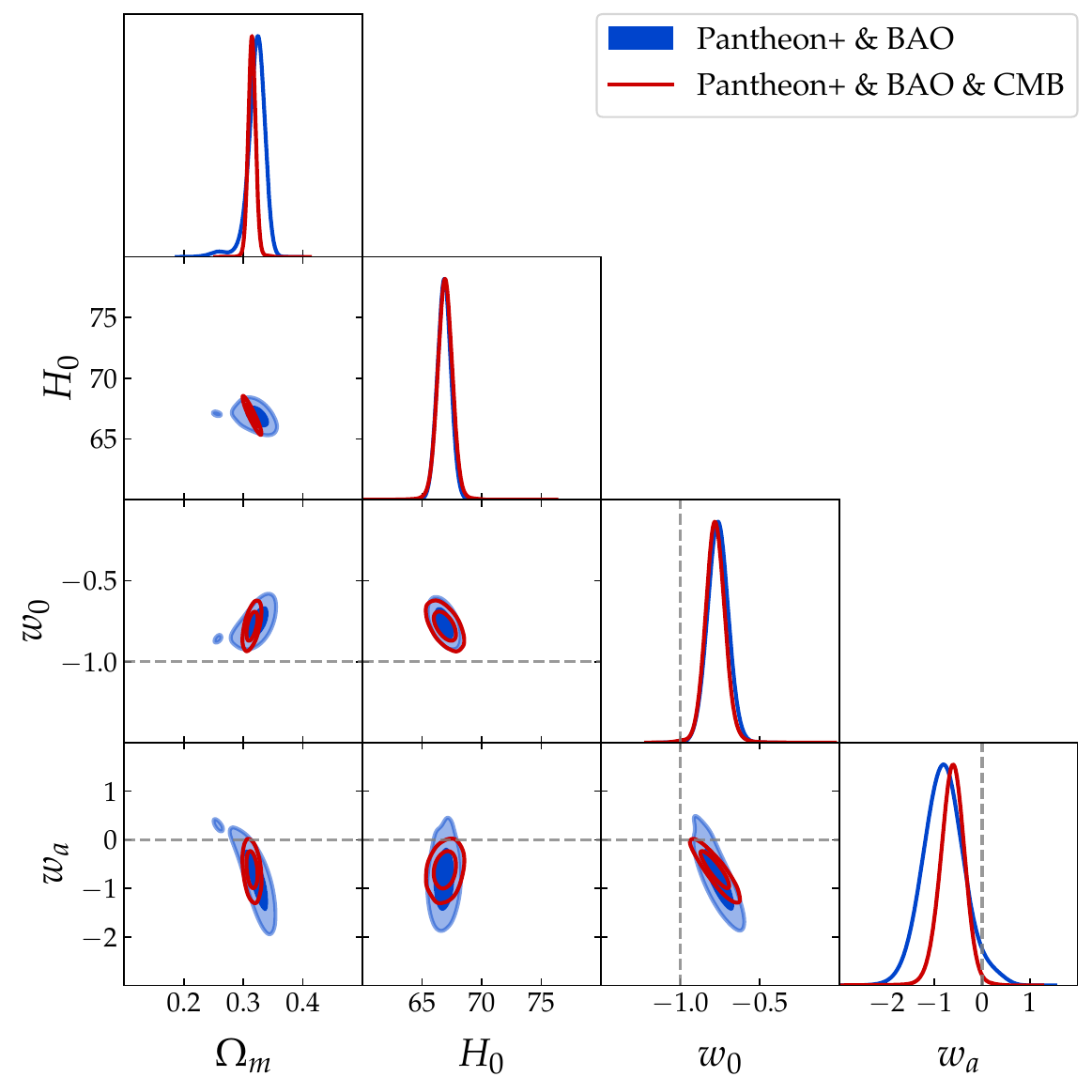}
    \includegraphics[width=0.49\linewidth]{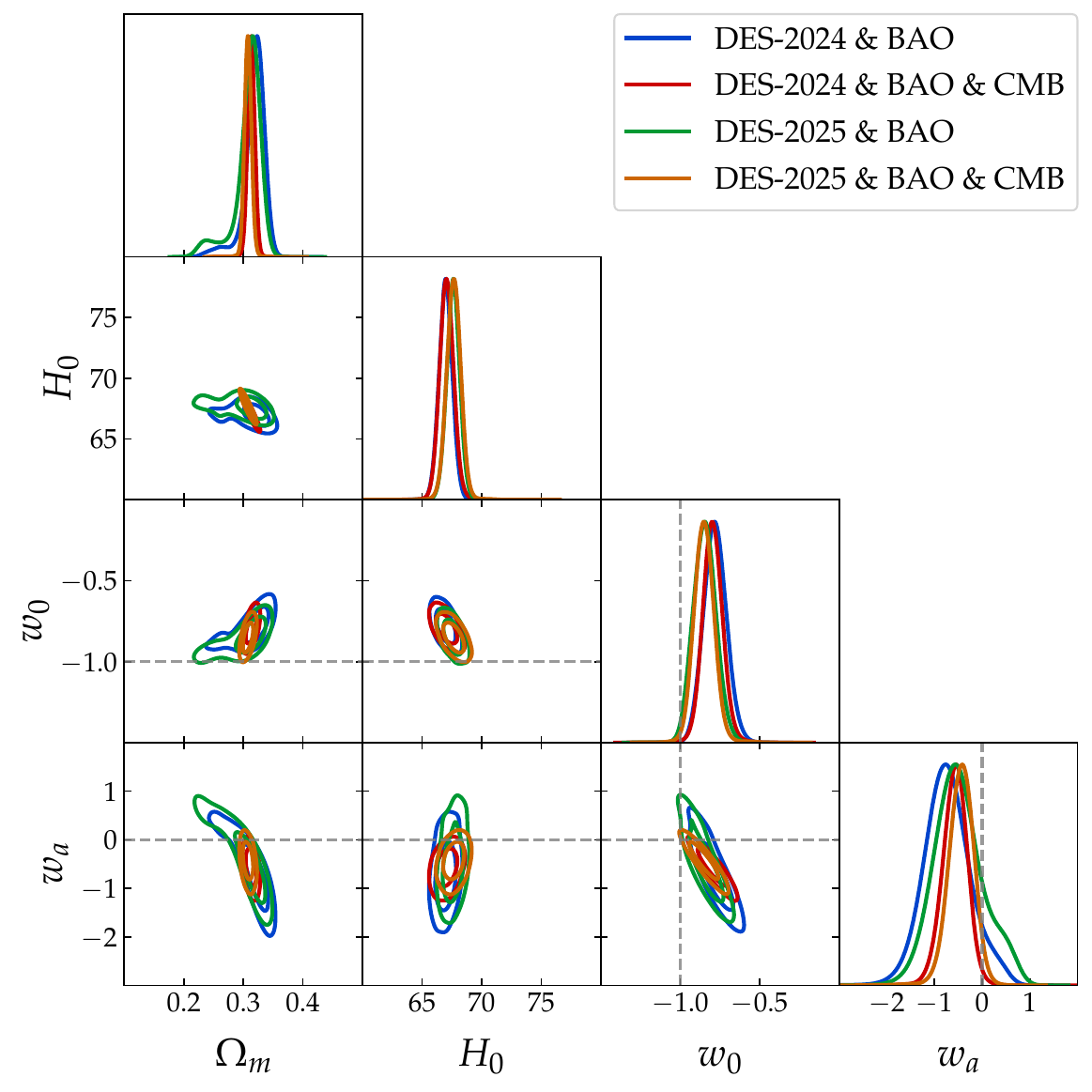}\\
    \makebox[0.49\linewidth]{(a)} \makebox[0.49\linewidth]{(b)}
    \caption{$w_0w_a$CDM posteriors without and with the Planck CMB prior, for  {(\textbf{a})} Pantheon$+$,  {(\textbf{b})}~DES-SN5YR~2024 and DES-SN5YR~2025, each combined with DESI~BAO.  Dashed lines mark the $\Lambda$CDM point $(w_0,w_a)=(-1,0)$.}
    \label{fig:cosmo_pairs}
\end{figure}

We now turn to the full $w_0w_a$CDM model and study how the dark-energy constraints depend on the supernova sample and on the inclusion of the CMB prior. We consider three supernova compilations (Pantheon$+$, DES-SN5YR 2024 and 2025), each combined with DESI BAO, and contrast the fit without and with the Planck prior.  The best-fit parameters of the six emulators are collected in Table~\ref{tab:cosmo_bestfit}, and the corresponding posteriors are compared pairwise in Figure~\ref{fig:cosmo_pairs}.

A consistent pattern emerges.  Without the CMB prior, all three supernova$+$BAO combinations prefer a dynamical dark energy with $w_0>-1$ and $w_a<0$, i.e.\ an equation of state that crosses the phantom divide as it evolves, the same hint of evolving dark energy recently reported by DESI.  The associated contours are broad, with $\sigma(w_a)\approx0.4$, reflecting the well-known $w_0$--$w_a$ geometric degeneracy that the surrogate captures as an elongated, curved ridge. The deviation is strongest for Pantheon$+$ and DES-2024 and milder for the DES-2025 update, which sits between the two.  Adding the Planck prior pulls every fit back towards $\Lambda$CDM ($w_0\approx-0.96$ to $-0.97$, $w_a\approx0.02$) and tightens the constraints dramatically: the uncertainty on $w_a$ shrinks by about a factor of ten (from $\sim0.42$ to $\sim0.045$), on $w_0$ by a factor of five, and on $\Omega_m$ by a factor of three.  Figure~\ref{fig:cosmo_pairs} makes this contraction visually explicit for the three samples.

\begin{table}[H]
 
 \caption{Best-fit parameters of the $w_0w_a$CDM emulators for the three supernova samples combined with DESI~BAO, without and with the Planck CMB prior. Each row is a separate XGBoost surrogate; $R^2$ is the validation score and $\sigma(w_a)$ the marginalised $1\,\sigma$ uncertainty on $w_a$.  Adding the CMB prior tightens the constraints moderately and shifts the fits partway towards $\Lambda$CDM, but a dynamical-dark-energy preference ($w_0>-1$, $w_a<0$) survives in every combination.}
\label{tab:cosmo_bestfit}

\setlength{\tabcolsep}{3.3mm}
\resizebox{\textwidth}{!}{\begin{tabular}{lcccccccc}
\toprule
\textbf{Data Set} & \boldmath$\Omega_m$ & \boldmath$H_0$ & \boldmath$w_0$ & \boldmath$w_a$ & \boldmath$\chi^2_{\rm min}$ & \boldmath$R^2$ & \boldmath$\sigma(w_a)$ \\
\midrule
Pantheon$+$ $\&$ BAO   & 0.323 & 66.9 & $-0.766$ & $-0.785$ & 1686.9 & 0.9978 & 0.42 \\
Pantheon$+$ $\&$ BAO $\&$ CMB & 0.314 & 66.9 & $-0.783$ & $-0.600$ & 1691.8 & 0.9994 & 0.24 \\
DES-2024 $\&$ BAO  & 0.322 & 67.0 & $-0.782$ & $-0.726$ & 1654.6 & 0.9976 & 0.45 \\
DES-2024 $\&$ BAO $\&$ CMB & 0.313 & 67.1 & $-0.801$ & $-0.542$ & 1659.3 & 0.9994 & 0.25 \\
DES-2025 $\&$ BAO  & 0.314 & 67.6 & $-0.843$ & $-0.536$ & 1647.0 & 0.9976 & 0.45 \\
DES-2025 $\&$ BAO $\&$ CMB    & 0.308 & 67.7 & $-0.855$ & $-0.405$ & 1651.3 & 0.9993 & 0.25 \\
\bottomrule
\end{tabular}}

\end{table}

\vspace{-6pt}

Reading the Pantheon$+$ panel directly (the DES panels are qualitatively identical), the supernova$+\allowbreak$BAO posterior (blue, filled) is a wide, tilted ellipse in the $w_0$--$w_a$ plane, elongated along the anti-correlation direction and displaced from the $\Lambda$CDM point (dashed lines) towards $w_0>-1$, $w_a<0$, whereas the $+$CMB posterior (red) collapses to a small blob centred on $(w_0,w_a)=(-1,0)$; the $\Omega_m$ and $H_0$ marginals, already well determined, barely move.  The supernova$+$BAO best fits, for instance $(w_0,w_a)\approx(-0.78,-0.73)$ for DES-2024, fall in the same region and direction of the $w_0$--$w_a$ plane as the dynamical-dark-energy preference reported by the DESI DR2 analysis for the analogous DESI$+$supernova combinations, so the emulator reproduces that published result.

The CMB enters here not as a forward computation of the temperature and polarisation spectra, but as a compressed Gaussian constraint built from the Planck PR4$+$lensing covariance matrix (Eq.~\ref{eq:chi2_cmb}), a standard and economical way to fold the CMB into a late-time fit.  It carries the Planck constraints on all four parameters, with the $H_0$--$\omega_m$ correlation retained, and in particular the tight bounds on the dark-energy sector that the supernovae and BAO leave nearly free; this is what allows it to tighten $w_0$ and $w_a$ by up to an order of magnitude and to draw the joint fit into agreement with $\Lambda$CDM.  The relevant point for the present work is that the surrogate reproduces this compressed-CMB likelihood as faithfully as the supernova$+$BAO one, so the contraction of the contours is a genuine output of the emulated fit and not an artefact of the emulation.

\subsubsection{Computational benchmark: surrogate versus exact likelihood}\label{sec:cosmo_bench}

\begin{figure}[H]
    
    \includegraphics[width=0.49\linewidth]{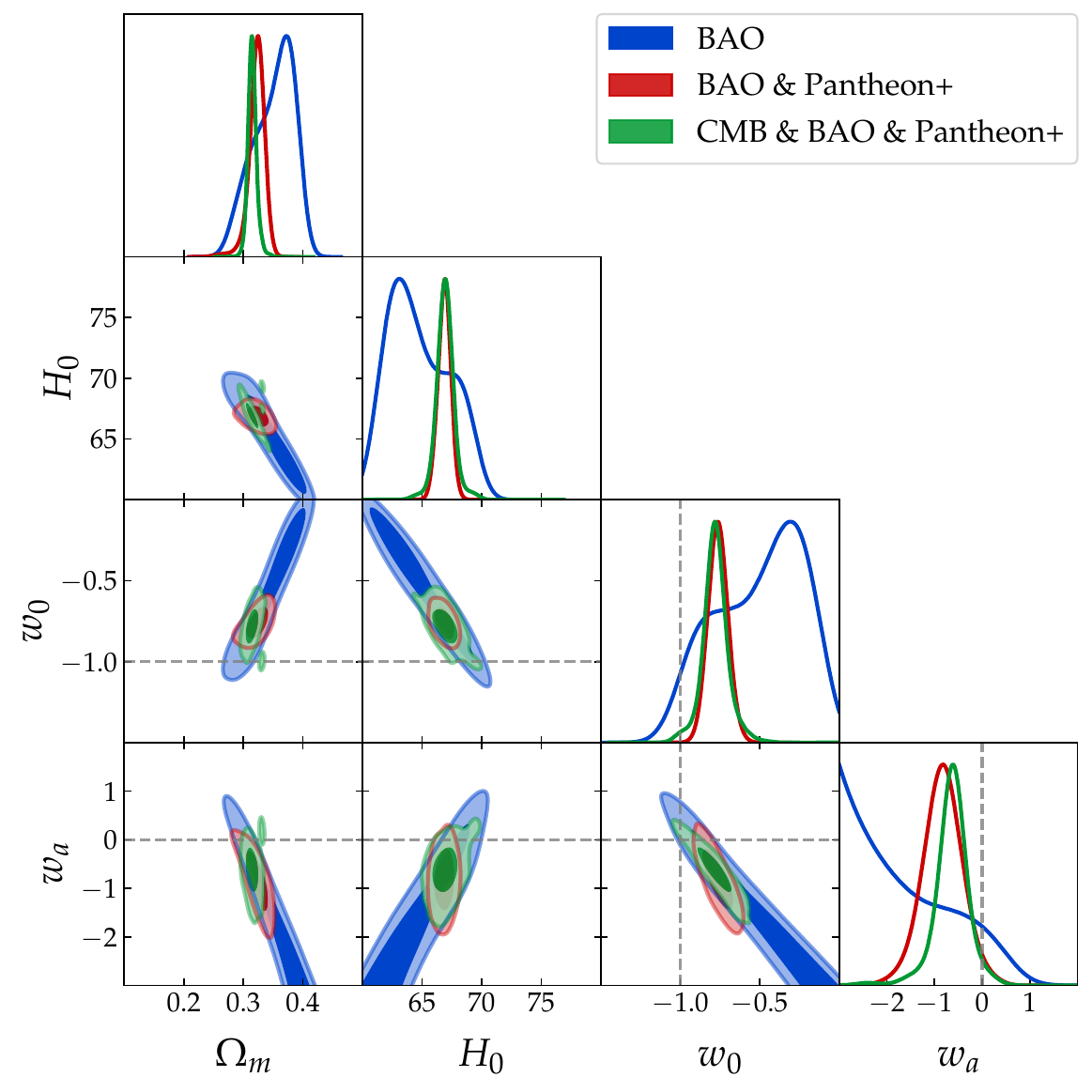}\hfill
    \includegraphics[width=0.49\linewidth]{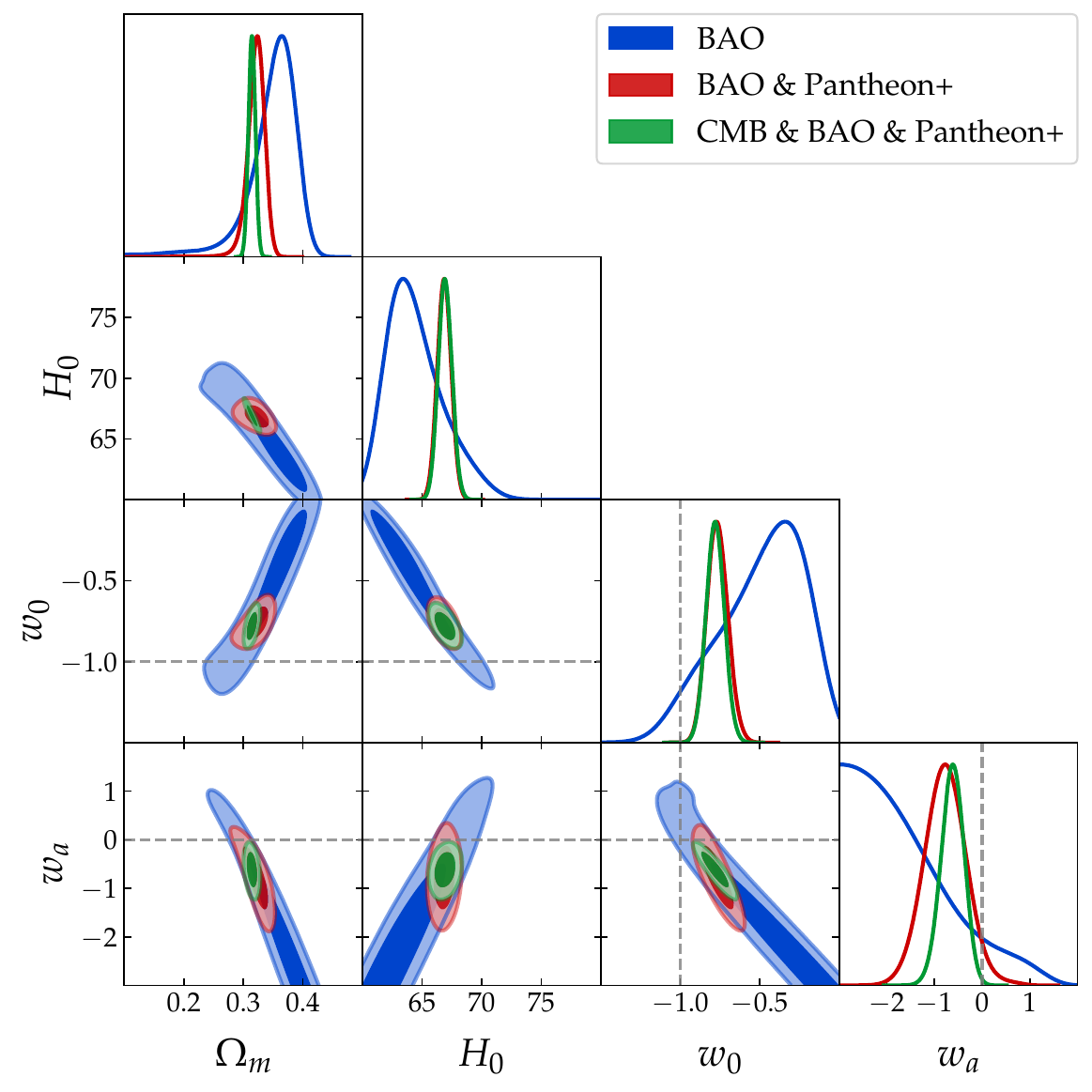}\\
    \makebox[0.49\linewidth]{(a)}\hfill\makebox[0.49\linewidth]{(b)}
    \caption{$w_0w_a$CDM posteriors for the three nested data combinations,
    BAO only (blue), BAO \& Pantheon$+$ (red) and CMB \& BAO \& Pantheon$+$ (green),
    sampled  {(\textbf{a})} from the XGBoost surrogate and  {(\textbf{b})} from the
    exact likelihood. The two are statistically indistinguishable,
    as are the four likelihood back-ends of Table~\ref{tab:cosmo_timing}.}
    \label{fig:cosmo_3way}
\end{figure}

The cosmological application also offers a clean setting to quantify the computational gain of the ML emulation, because here the exact likelihood is cheap enough to be sampled directly for comparison, unlike the flavour case, where a brute-force scan is prohibitive.  We benchmark three nested data combinations of increasing constraining power, BAO only, BAO \& Pantheon$+$ and CMB \& BAO \& Pantheon$+$, which illustrate how each probe closes the $w_0$--$w_a$ degeneracy: with only the 13 BAO points the four-parameter fit is essentially unconstrained ($\chi^2_{\rm min}=5.6$, with $w_a$ railing against the prior edge), adding the supernovae closes the contour, and the CMB prior tightens it onto $\Lambda$CDM (Figure~\ref{fig:cosmo_3way}).

We run the same random-walk Metropolis--Hastings sampler (1024 chains, identical settings and seed) with four interchangeable likelihood back-ends: the XGBoost surrogate evaluated on the GPU and the exact $\chi^2$ computed with (i)~the \texttt{astropy}-based reference parallelised over a CPU process pool (``Astropy''), (ii)~a vectorised NumPy reimplementation of the distance integrals on the CPU (``CPU''), and (iii)~a \texttt{JAX} \texttt{vmap} kernel on the GPU (``GPU'')~\cite{jax2018github}.  All four back-ends are run on the same workstation, an 18-core/36-thread Intel Core i9-10980XE CPU ($3.0$\,GHz base, $4.8$\,GHz turbo) with $188$\,GB of RAM and an NVIDIA RTX~A6000 GPU ($48$\,GB). The Astropy reference must be evaluated one cosmology at a time and cannot be vectorised across the sampled points; its only parallelism is to spread those points over a process pool (up to 16 workers), which scales poorly and becomes impractical for the large supernova likelihoods (the n/a entries in Table~\ref{tab:cosmo_timing}). This is precisely why the distance integrals are reimplemented directly as the vectorised NumPy (CPU) and JAX (GPU) back-ends, which evaluate the whole batch of points in a single call.  Crucially, all four back-ends produce statistically identical posteriors, confirming that the surrogate is accurate and not merely fast.  The wall-clock times are summarised in Table~\ref{tab:cosmo_timing}: the surrogate completes each run in $4.8$--$7.4$\,s, $3.6$--$22.4\times$ faster than the CPU (vectorised NumPy) likelihood and $7.2$--$12.1\times$ faster than the exact GPU kernel, with the advantage growing for the more expensive Pantheon$+$ likelihoods.
\begin{table}[H]
 \caption{Wall-clock time of the posterior exploration (1024 chains, identical sampler settings) for the three nested data combinations, comparing the XGBoost surrogate with the exact likelihood evaluated using the \texttt{astropy} reference (Astropy, process pool), a vectorised NumPy kernel (CPU) and \texttt{JAX} on the GPU (GPU). The last two columns give the speed-up of the surrogate over the exact CPU and GPU likelihoods. The Astropy process pool is reported only for the BAO-only case (it was impractical for the large supernova likelihoods).}
\label{tab:cosmo_timing}

\setlength{\tabcolsep}{3.3mm}
\resizebox{\textwidth}{!}{\begin{tabular}{lcccccc}
\toprule
\multirow{2}{*}{\textbf{Data Set}}
& \multirow{2}{*}{\textbf{ML}} & \multicolumn{3}{c}{\textbf{Exact \boldmath$\chi^2$}} & \multicolumn{2}{c}{\textbf{Speed-Up}} \\
 &  & \textbf{Astropy} & \textbf{CPU} & \textbf{GPU} & \textbf{\boldmath$\times$CPU} & \textbf{\boldmath$\times$GPU} \\
\midrule
BAO only  & 7.4\,s & 98.5\,s & 26.9\,s  & 89.9\,s & 3.6  & 12.1 \\
BAO \& Pantheon$+$  & 6.8\,s & n/a
    & 151.4\,s & 82.1\,s & 22.4 & 12.1 \\
CMB \& BAO \& Pantheon$+$ & 4.8\,s & n/a  & 98.6\,s  & 34.3\,s & 20.8 & 7.2  \\
\bottomrule
\end{tabular}}

\end{table}
The structure of Table~\ref{tab:cosmo_timing} reflects two competing effects. The emulator's per-evaluation cost is essentially independent of the dataset, since a forward pass through the tree ensemble does not depend on the number of observables, whereas the exact likelihood scales with the data; this is why the speed-up over the CPU back-end grows from $\sim\!4\times$ for the 13~BAO points to $\gtrsim\!20\times$ for the $\sim\!1700$-supernova Pantheon$+$ fits. The exact GPU kernel behaves differently at small data volumes: for BAO only, its fixed kernel-launch and host--device-transfer overhead dominates the small per-step computation, so it is slower than the CPU (vectorised NumPy) code ($90$ versus $27$\,s), whereas for the much heavier Pantheon$+$ likelihood that overhead is amortised and the GPU is faster. Finally, because every chain is run to a fixed effective-sample-size target rather than a fixed number of steps, the wall-clock time also reflects how quickly each posterior mixes: the tightly CMB-constrained fit converges in fewer steps, so BAO \& Pantheon$+$ \& CMB is less expensive than BAO \& Pantheon$+$ despite containing more data.

The same acceleration benefits the up-front cost of building the training set.  Evaluating the $2\times10^{5}$ exact $\chi^2$ points needed to train a supervised surrogate takes about $4\times10^{3}$\,s (more than an hour) with the \texttt{astropy} reference on 16 CPU cores for the Pantheon$+$ combinations, whereas the batched \texttt{JAX}/GPU implementation of the distance integrals produces the same set in a few seconds. The GPU and Astropy evaluations agree to better than one unit in $\chi^2$ near the minimum, where the training signal matters; they differ only in the extreme tails ($\chi^2\!\sim\!10^{6}$), a harmless discrepancy since those points sit in the compressed part of the shifted-log transform of Equation~(\ref{eq:log_transform}). Thus both the construction and the sampling of the emulated likelihood are accelerated, and the one-time training cost is amortised over the whole posterior analysis, after which the surrogate replaces the physics computation~entirely.

\subsubsection{Feature importance}\label{sec:cosmo_shap}

\begin{figure}[H]
     \centering
    \includegraphics[width=0.49\linewidth]{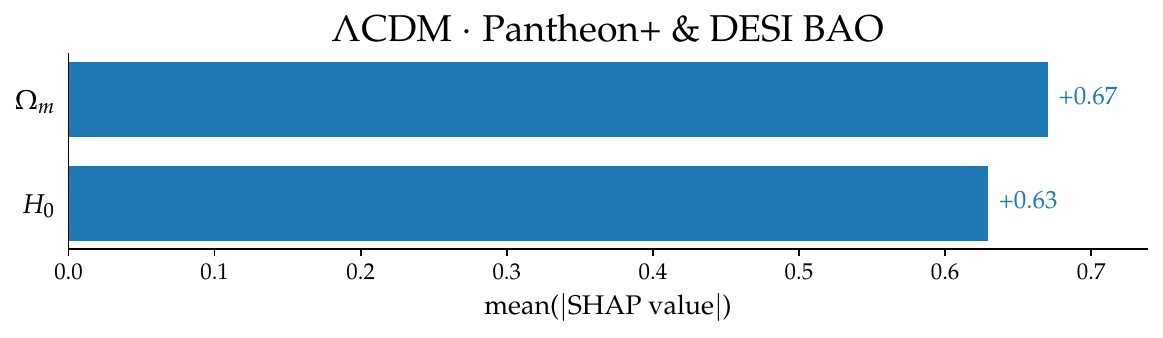}\\
    \makebox[0.49\linewidth]{(a)}\\[1ex]
    \includegraphics[width=0.49\linewidth]{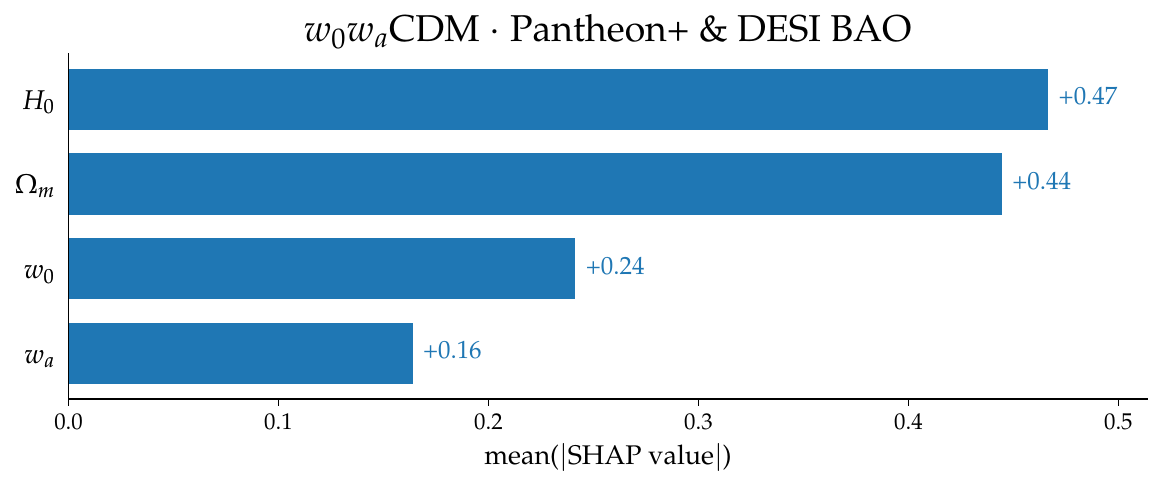}\hfill
    \includegraphics[width=0.49\linewidth]{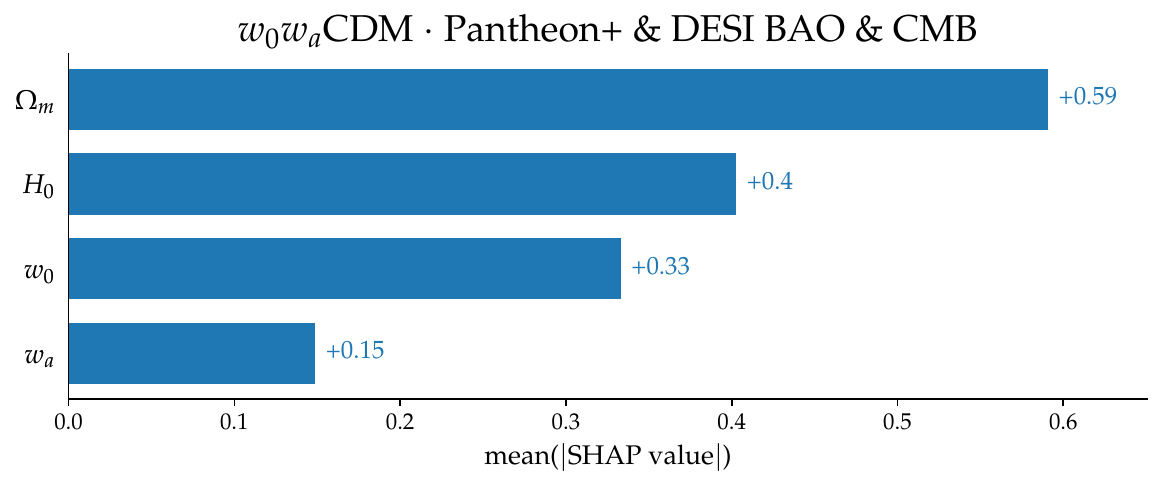}\\
    \makebox[0.49\linewidth]{(b)}\hfill\makebox[0.49\linewidth]{(c)}
    \caption{Global SHAP feature importance (mean absolute SHAP value, i.e.,\ each parameter's average contribution to the emulated $\chi^2$) for  {(\textbf{a})} the $\Lambda$CDM baseline,  {(\textbf{b})} $w_0w_a$CDM with Pantheon$+$ \& DESI~BAO, and  {(\textbf{c})} the same fit with the CMB prior
    added.}
    \label{fig:cosmo_shap}
\end{figure}

As in the flavour analysis, SHAP values provide a transparent, model-independent ranking of the parameters by their mean absolute contribution to the emulated $\chi^2$, that is, by how steeply the likelihood responds to each one across the sampled volume, which is a direct proxy for how tightly each parameter is constrained. The rankings (Figure~\ref{fig:cosmo_shap}) follow the expected physics and, tellingly, reorganise when the CMB prior is added.

In the $\Lambda$CDM baseline the two parameters are almost tied, $\Omega_m$ ($0.67$) just above $H_0$ ($0.63$): both are comparably and tightly determined by the joint supernova$+$BAO geometry, with $\Omega_m$ marginally ahead because it enters every distance through $E(z)$.  In the $w_0w_a$CDM supernova$+$BAO fit the background parameters dominate, $H_0$ ($0.47$) and $\Omega_m$ ($0.44$) well above the dark-energy shape parameters $w_0$ ($0.24$) and $w_a$ ($0.16$).  Within the dark-energy sector $w_0$ always ranks above $w_a$, and this ordering has a clear physical origin: $w_0$ is the value of the equation of state today, so it controls $w(z)$ at low redshift, exactly where dark energy dominates the energy budget and where the supernovae and BAO are most numerous and constraining, whereas $w_a$ only sets the evolution through the $z/(1+z)$ factor, which vanishes at $z=0$ and grows only towards higher redshift, where dark energy is subdominant to matter and the data are sparser. $w_0$ therefore has far more leverage on the observables than $w_a$, as the marginalised errors confirm ($\sigma(w_0)\approx0.066$ versus $\sigma(w_a)\approx0.42$, i.e.\ $w_0$ about six times better determined), and $w_a$, the flattest and most degenerate direction, ranks last; this is the feature-importance counterpart of the elongated $w_0$--$w_a$ ridge of Figure~\ref{fig:cosmo_pairs}.  Adding the compressed CMB prior inverts the background-versus-dark-energy hierarchy: $w_0$ jumps to the top ($0.53$) and $w_a$ to second ($0.27$), while $\Omega_m$ ($0.25$) and $H_0$ ($0.24$) drop.  The CMB prior adds tight, direct constraints on $w_0$ and $w_a$, which the supernovae and BAO left almost free, so the $\chi^2$ now rises steeply along $w_0$ and $w_a$ and they take most of the SHAP weight, whereas $\Omega_m$ and $H_0$, already well measured and consistent with the prior, contribute proportionally less. Note that $w_0$ stays ahead of $w_a$ throughout (the same $\sim\!2\!:\!1$ ratio with and without the prior), for the same low-redshift-leverage reason. The SHAP reordering is thus the feature-level fingerprint of the contour contraction seen in Figure~\ref{fig:cosmo_pairs}, and the agreement between this data-driven ranking and the expected constraining structure confirms that the surrogate has learned the correct physical dependencies rather than numerical artefacts.

\begin{figure}[H]
    \centering
    \includegraphics[width=0.62\linewidth]{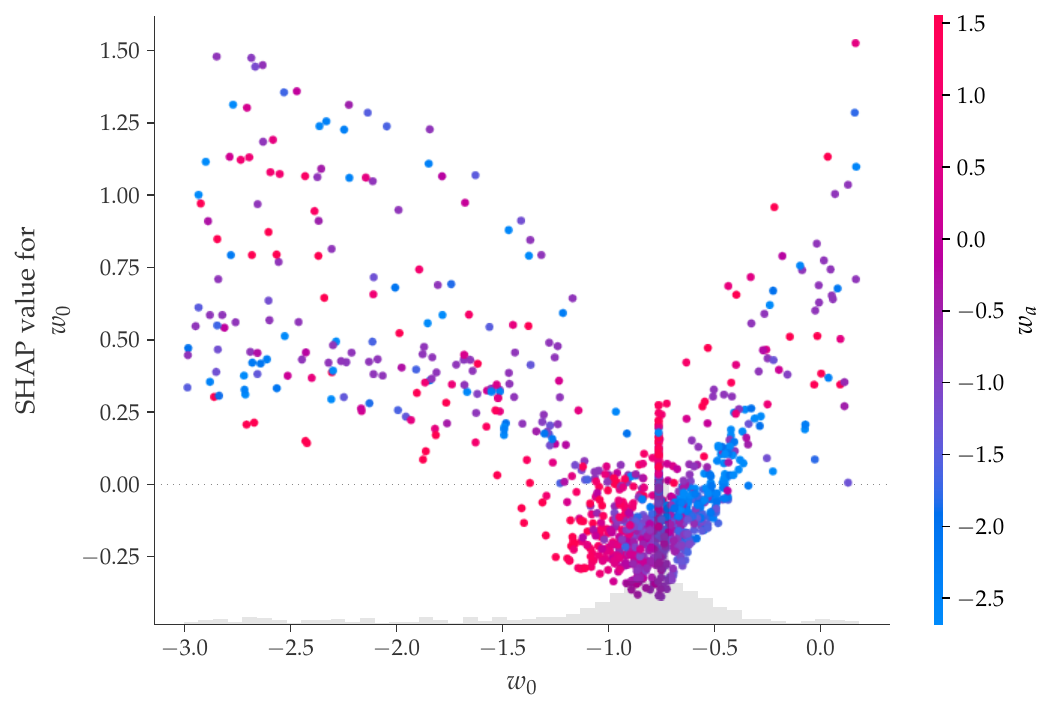}
    \caption{Point-wise SHAP values for $w_0$ in the $w_0w_a$CDM fit to Pantheon$+$\,$\&$\,DESI~BAO.  Each point is a sampled cosmology: its horizontal position is the value of $w_0$, its vertical position the SHAP value of $w_0$ (its signed contribution to the emulated $\chi^2$ at that point), and its colour the value of $w_a$.}
    \label{fig:cosmo_shap_w0}
\end{figure}

Beyond this global ranking, SHAP also resolves how each parameter acts point by point across the sampled volume. Figure~\ref{fig:cosmo_shap_w0} shows this for $w_0$ in the Pantheon$+$\,$\&$\,DESI~BAO fit: its SHAP value, the signed contribution of $w_0$ to the emulated $\chi^2$ at each sampled cosmology, plotted against $w_0$ and coloured by $w_a$.  The points trace a clear parabola, the data-driven image of the $\chi^2$ well along $w_0$: the contribution is least, and slightly negative, in the densely sampled best-fit region near $w_0\!\approx\!-0.8$, where $w_0$ sits at its optimum and helps lower the $\chi^2$, and grows steeply and positive as $w_0$ departs in either direction and drives the $\chi^2$ up.  It is the magnitude of this contribution, not its sign, that defines importance: the mean absolute value reported in Figure~\ref{fig:cosmo_shap} is dominated by these steep flanks, so the parabola itself is the fingerprint of a tightly constrained parameter, whereas a flat, weakly constrained direction such as $w_a$ shows small SHAP values throughout and no comparable well. The small contribution in the best-fit cloud therefore does not mean that $w_0$ is unimportant there; it simply reflects that, at the bottom of the well, the likelihood is locally flat in $w_0$, exactly as expected for a parameter pinned at its optimum. The colour finally encodes the residual $w_0$--$w_a$ interaction: at fixed $w_0$ the contribution still varies with $w_a$, the feature-level counterpart of the elongated $w_0$--$w_a$ degeneracy of Figure~\ref{fig:cosmo_pairs}.  Analogous per-parameter plots for the remaining parameters and data combinations are collected in the public \href{https://github.com/AlejandroMirRamos/PhysicsML}{\texttt{PhysicsML}} repository (see the Data Availability Statement).

The cosmological application plays a double role in this work.  Methodologically it is a validation of the ML strategy: unlike the flavour and ALP likelihoods, the exact cosmological $\chi^2$ is cheap enough to be sampled directly, so we could run the same MCMC with four independent back-ends (Section~\ref{sec:cosmo_bench}) and verify that the surrogate posteriors are statistically identical to the exact ones.  This is a check that cannot be carried out where the exact likelihood is prohibitive, and it certifies that the order-of-magnitude speedup of the emulator does not come at the price of accuracy.

Beyond this cross-check, the cosmological fits make the practical advantages of the approach concrete.  First, speed: once trained, the emulator samples each posterior in a few seconds against tens to hundreds of seconds for the exact likelihood, and even the up-front construction of the training set is accelerated on the GPU, so the cost is paid once and amortised over the whole analysis; the advantage grows with the cost of the underlying likelihood, which is precisely why it becomes decisive in the flavour and ALP applications. Second, robustness and adaptability: a single pipeline handles the two-parameter $\Lambda$CDM fit and the four-parameter $w_0w_a$CDM one, three different supernova samples, the presence or absence of the CMB prior, and likelihoods that are tightly ($\chi^2_{\rm min}\!\approx\!1700$) or barely ($\chi^2_{\rm min}\!\approx \!5$) constrained, with only the shifted-log target of Eq.~(\ref{eq:log_transform}) needed to make the surrogate resolve all of them. Third, resolution: the emulator returns smooth, closed confidence regions even for the strongly curved $w_0$--$w_a$ degeneracy and the nearly flat BAO-only landscape, where a direct grid would be either too coarse or too expensive. Fourth, transparency: the SHAP analysis recovers the physically expected constraining hierarchy and its reorganisation when the CMB is added, so the gain in speed does not turn the inference into a black box.  Taken together, these results show that the framework of Section~\ref{sec:ml} transfers cleanly from high-energy physics to cosmology, the cosmological case providing the controlled benchmark that certifies the method while the more expensive applications reap the largest computational gains.

\section{Conclusions}\label{sec:conclusions}

We have presented a general and interpretable machine-learning framework for accelerating likelihood-based inference in high-energy physics and cosmology. Using XGBoost surrogate models combined with SHAP interpretability and parallel Markov chain Monte Carlo sampling, the framework efficiently explores high-dimensional, non-Gaussian parameter spaces. This demonstrates that the approach is general rather than tied to a particular physical system.

The surrogate's robustness across diverse likelihoods stems from three core methodological choices. First, the training design combines a global space-filling prior sample with a local, Hessian-covariance-scaled best-fit cloud to automatically resolve narrow, anisotropic degeneracies. This design is augmented by an active-learning Gaussian process expected to offer improvement in sparse or sharply structured regions. Second, training targets are dynamically adapted to the local landscape, using a direct log-likelihood for the flavour fit, a two-stage classifier-plus-regressor via a sigmoid transform for the axion-like-particle fit, and a shifted-$\log_{10}$ transform to resolve cosmological confidence regions across a $\chi^2_{\rm min}$ spanning two orders of magnitude. Finally, regression trees partition the parameter space adaptively to capture narrow ridges, sharp transitions, and plateau-like regions without neural-network oversmoothing. The space-filling baseline ensures that emulator evaluations remain strictly within the validated training domain.

Across the three cases studied, the framework delivered substantial computational gains while closely tracking the exact likelihood. We distinguish two levels of this statement: the surrogate is robustly reliable as an accelerator for exploratory scans and adaptive sampling, while its use as a statistically faithful replacement for the exact likelihood in the low-$\Delta\chi^2$ region that sets the confidence limits is validated explicitly in all three applications. This validation is most stringent in the cosmological case, where the exact likelihood can be sampled directly. Likelihood evaluations were accelerated by factors of up to $5\times10^5$ in the flavour and axion-like-particle analyses, while posterior sampling in the cosmological application achieved speed-ups ranging from $4$ to $22$ over the CPU (vectorised NumPy) implementation and from $7$ to $12$ over the JAX/GPU kernel. The advantage grows as the underlying likelihood becomes more expensive. A further key advantage of the framework is its interpretability. SHAP analyses consistently recovered the expected hierarchy of parameter and observable importance, providing a transparent physical interpretation of the surrogate predictions. The cosmological case also provides a controlled validation: because the exact likelihood could be sampled with four independent back-ends, we could verify directly that the surrogate posteriors are statistically identical to the exact ones.

The three applications show that the framework reproduces the relevant phenomenology at a fraction of the computational cost. It demonstrates that the almost-perfect $R_D$--$\mathrm{BR}(B^+\to K^+\nu\bar\nu)$ correlation found in the previous Scenario~II~\cite{Alda:2021rgt} fit disappears when $C_1$ and $C_3$ vary independently, identifies the axion-like-particle parameter region favoured by the Belle~II $B^+\to K^+\nu\bar\nu$ excess, and recovers the preference for dynamical dark energy in supernova$+$BAO data, which persists at the $2.5$ to $3.8\,\sigma$ level once the compressed Planck CMB prior is included. In all cases, SHAP values transparently identify the measurements driving the results.

Overall, our results demonstrate that interpretable Machine-Learning surrogates provide a fast, accurate, and scalable solution for likelihood-based inference. The framework is readily transferable to other global-fit problems and offers a promising foundation for future developments, including automated active learning, differentiable emulators, and applications to higher-dimensional parameter spaces.

\section*{Acknowledgments}
This work is partially supported by Spanish MINECO/FEDER Grants PGC2022-126078NB-C21 and PID2024-160228NB-I00, funded by MCIN/AEI/10.13039/\ 501100011033 and “ERDF A way of making Europe”, and Grant E21-23R funded by Arag\'on Government and the European Union, and the NextGenerationEU Recovery and Resilience Program on Astrof\'isica y F\'isica de Altas Energ\'ias CEFCA-CAPA-ITAINNOVA. Jorge Alda thanks the warm hospitality of University of Zaragoza/CAPA and of the Kavli Institute for the Physics and Mathematics of the Universe in Tokyo during the completion of this work.\\
{\em{Data availability}}: The code, Jupyter notebooks and data used to produce the results of this paper are publicly available in the \href{https://github.com/AlejandroMirRamos/PhysicsML}{\texttt{PhysicsML}} repository, which gathers the three applications presented here (semileptonic $B$ meson anomalies, axion-like particles, and cosmology). Each subproject provides the full pipeline used to build the training sets, train the XGBoost surrogate, compute the SHAP analysis and sample the posterior with MCMC, together with the input experimental data required to reproduce the figures and tables.

\bibliographystyle{utphys.bst}
\bibliography{biblio}

\end{document}